\documentclass[11pt]{article}

\usepackage{epsfig}
\usepackage{amsmath,amssymb}
\usepackage{mathtools}
\usepackage{braket}
\usepackage{hyperref}
\usepackage{authblk}
\usepackage{graphicx}
\usepackage{comment}
\usepackage{soul}
\usepackage{xspace}
\graphicspath{ {./images/} }
\usepackage[section]{placeins}
\usepackage{color}
\usepackage{subcaption}

\begin{document}
\date{}

\begin{titlepage}
\begin{center}

{\Large Near-horizon modifications in finite $N$ holography}

\vspace{12mm}

\renewcommand\thefootnote{\mbox{$\fnsymbol{footnote}$}}
Rishkrith Bairy${}^{1,2}$\footnote{25d1084@iitb.ac.in},
Mrityunjay Nath${}^{1}$\footnote{nath.mrityunjay@gmail.com}, and
Debajyoti Sarkar${}^{1}$\footnote{dsarkar@iiti.ac.in}

\vspace{6mm}

\vspace{2mm}
${}^1${\small \sl Department of Physics} \\
{\small \sl Indian Institute of Technology Indore} \\
{\small \sl Khandwa Road 453552 Indore, India}

\vspace{4mm}
${}^2${\small \sl Department of Physics} \\
{\small \sl Indian Institute of Technology Bombay} \\
{\small \sl Powai, Mumbai 400076, India}

\end{center}

\vspace{12mm}

\noindent

If one extends the AdS/CFT extrapolate dictionary to large but finite $N$, we are expected to obtain non-perturbative violations of bulk semi-classicality and micro-causality. 
Previously this was achieved by implementing a late boundary time cut-off, while smearing the boundary operator via the HKLL kernel.
By performing explicit bulk reconstructions in the backgrounds of near-horizon modified AdS$_2$ and BTZ black holes, we recover similar deviations from semi-classicality as above. For these black hole mimickers, the near-horizon modification is controlled by a throat parameter which sets the scale of this deviation.
In three bulk dimensions, probe dynamics, for a fixed radial band, also exhibits a dip-ramp-plateau structure in their spectral form factor when averaged over the throat parameter. Such structure has also been found recently in the background with a stretched horizon or a brick wall.

\end{titlepage}
\setcounter{footnote}{0}
\renewcommand\thefootnote{\mbox{\arabic{footnote}}}

\hrule
\tableofcontents
\bigskip
\hrule

\addtolength{\parskip}{8pt}

\newpage

\section{Introduction}\label{sec:intro}

One of the important goals of the Anti de Sitter (AdS$_{d+1}$)/ Conformal field theory (CFT$_d$) correspondence \cite{Maldacena:1997re,Gubser:1998bc,Witten:1998qj} is to understand how far we can deform the large central charge (often denoted by parameter $N$ in the literature) version of the duality, and what form does the correspondence take when we look at its finite $N$ limit. There has been an enormous amount of progress in this direction since the inception of AdS/CFT itself, more recently along the lines of operator algebras in the bulk and the boundary (for a recent review see \cite{Liu:2025krl}, although the present work will not focus along this line). On one hand, at finite $N$, the CFT Hilbert space is effectively finite in a microcanonical window, leading to discrete spectra and late-time unitarity features in boundary observables \cite{Barbon:2003aq,Barbon:2014rma,Dyson:2002nt,Maldacena:2001kr}. However, on the other hand, these features are in apparent tension with the emergence of semiclassical locality in the bulk at the large $N$ limit \cite{Hamilton:2005ju,Hamilton:2006az,Hamilton:2006fh}. We will call these local bulk operators HKLL fields (as they are commonly known in the literature), and the modifications due to the above finite $N$ physics are often expected to appear through non-perturbative corrections to HKLL fields \cite{Hamilton:2007wj,Kabat:2014kfa}.

In fact, in our current work, we will probe the finite $N$ effects mostly through the perspectives of bulk reconstruction (in the spirit of HKLL) methods, and via the studies of spectral form factor, that appears in the physics of quantum chaotic systems.  In the ordinary large $N$ HKLL reconstruction,\footnote{Here by large $N$, we will mean either infinite $N$, or perturbative $1/N$ corrections. Non-perturbative corrections will be discussed separately.} at various perturbative orders in $1/N$, bulk fields are expressed as smeared boundary operators supported on a boundary spacetime interval. The size of this interval is controlled by the bulk point and the background geometry. Additional subtleties arise in geometries with horizons. For example, in black hole backgrounds, the relevant geometric quantities (e.g.~tortoise/optical depth) can become unbounded near the horizon \cite{Hamilton:2007wj}. This suggests that naive reconstruction can call upon arbitrarily late boundary times. This is problematic. In a field theory system with finite entropy $N\sim S$ and inverse temperature $\beta$, the boundary correlators (of operators with conformal dimension $\Delta$) start showing small fluctuations of $\mathcal{O}(e^{-S/2})$ at time scale of order $\sim\frac{\beta S}{4\pi \Delta}$ (see e.g.~figure 2 of \cite{Kabat:2014kfa}).\footnote{These small fluctuations become $\mathcal{O}(1)$ at Heisenberg timescale $\sim e^S$. Poincar\'{e} recurrences happen in doubly exponential timescales, and they are rarely relevant for these discussions.} Following \cite{Kabat:2014kfa}, we have denoted this timescale as $t_{\rm max}$. Even though the timescale is itself of order $\sim S$ and may look perturbative, when these fluctuating correlators are integrated against the HKLL smearing functions, they lead to diverging integrals. This leads to a breakdown of the usual bulk reconstruction procedure. One of the proposed effective treatments is to therefore introduce an \emph{ad-hoc} late-time cut-off in the smearing integral. As we will see here, such a cut-off can also be interpreted as an ``excision'' of near-horizon region in the bulk, and they lead to exponentially suppressed departure from the semiclassical locality of the bulk fields \cite{Kabat:2014kfa}. Similar effects will also appear for a uniformly accelerated Rindler observer, which suggests that the non-locality may be observer dependent. 

In fact as a possible model of the excised spacetime, we will show that the Damour--Solodukhin (DS) type wormholes \cite{Damour:2007ap} provide a geometric realization of such finite $N$ motivated modifications. In these geometries a would-be horizon is replaced by a smooth throat controlled by a parameter $\lambda^2\sim e^{-cS}$. In the bulk, $S$ would have been the entropy of the semiclassical black hole, and $c$ is some constant depending on the bulk field contents (and therefore the conformal dimensions of the corresponding boundary operators), and the dimension of the spacetime.\footnote{For example, in the four dimensional analysis of \cite{Berthiere:2017tms}, $c$ is given by $\frac{1}{8\pi^2\bar{a}}$ for a very specific bulk field content. See e.g. equation (20) of that paper.} As a consequence of such a near-horizon modification, the range of the tortoise/optical coordinate becomes \emph{finite}. This finiteness automatically bounds the support of smearing integrals and therefore implements a geometric cut-off. This provides a natural, bulk motivation of the excision process implemented by \cite{Kabat:2014kfa}.

The paper is organized as follows. In section \ref{sec:AdS2_cutoff_excision}, we start by reviewing the repercussions of finite $N$ CFT excision procedure, which generically results into non-local bulk fields. In section \ref{sec:AdS2_DSW_cutoff} we then recover identical non-localities by taking the alternate perspective of semiclassical bulk dynamics in the two-dimensional DS wormholes. We then generalize our analysis to three dimensions in section \ref{sec:3Dheun}, where one can show that the resulting bulk field has $\mathcal{O}(\lambda^2)$ corrections compared to the semiclassical BTZ field, even though the usual bulk reconstruction becomes technically challenging. However, one may investigate the spectral form factor using WKB analysis, that we perform in section \ref{sec:numerics_SFF_3D}. This leads to a dip-ramp-plateau behavior in the spectral form factor of the probe dynamics. Finally, we make some concluding remarks and discuss some potential future directions in section \ref{sec:discussion_outlook}. The various appendices provide some further details of the calculations that were left out during the main text of the paper.

\section{Finite $N$ bulk reconstruction in AdS$_2$: cut-off and excision}
\label{sec:AdS2_cutoff_excision}

A concrete entry point to study the finite $N$ modified HKLL reconstruction is in AdS$_2$ Rindler. This was the main example of \cite{Kabat:2014kfa}, and this section can be considered as an unavoidable review of the procedure. Initially, we will be interested in massless bulk fields dual to a boundary scalar primary ${\cal O}$ of conformal dimension $\Delta=d=1$. The metric reads
\begin{equation}
    ds^2=-\frac{r^2-R^2}{R^2}dt^2+\frac{R^2}{r^2-R^2}dr^2\,,
\end{equation}\label{eq:AdS_2rindler}
where \(R\) is the AdS radius and the Rindler horizon is at \(r_h=R\). In the thermofield double (TFD) formalism, the two-point function of (right-) boundary operators is given by\footnote{Only when $L$ or $R$ appears in the subscript of an operator $\mathcal{O}$, it is to be interpreted as left or right CFT operator.}
\begin{equation}
\langle {\cal O}_R(t)\,{\cal O}_R(t')\rangle=\frac{1}{2R^2\bigl(1-\cosh\left(\frac{t-t'}{R}\right)\bigr)}\,.
\label{eq:OROR_intro}
\end{equation}
These correlators are essentially the CFT inputs to the standard HKLL reconstruction. If we are interested in a massless bulk scalar $\phi$ located in the exterior region $(r>R)$, then the resulting expression is written as a smearing integral over a \emph{finite} interval
\begin{equation}
\phi(t,r)=\frac{1}{2}\int_{t-\delta t}^{t+\delta t}dt'\,{\cal O}_R(t')
\qquad
\text{with}
\qquad
\delta t=\frac{R}{2}\log\frac{r+R}{r-R}\,.
\label{eq:HKLL_AdS2_intro}
\end{equation}
$\delta t$ here is essentially the tortoise or optical length defined as $\delta t=-\int\frac{R^2}{r^2-R^2}\,dr$. The key point is that even though $\delta t$ is finite for $r\gg R$, it grows without bound for bulk fields approaching the black hole horizon. This forces the smearing integral to access arbitrarily large boundary-time separations in computing bulk two-point functions via HKLL method. The regulated prescription introduced below is a proxy for the finite $N$ expectation that such arbitrarily large separations are not captured by semiclassical correlators.

Substituting \eqref{eq:HKLL_AdS2_intro} into \eqref{eq:OROR_intro}, one obtains a bulk to boundary two-point function given by 
\begin{equation}
\langle \phi(t,r)\,{\cal O}_R(t'')\rangle 
=\frac{1}{2}\int_{t-\delta t}^{t+\delta t}dt'\,\langle {\cal O}_R(t'){\cal O}_R(t'')\rangle\,,
\label{eq:phi_OR_start}
\end{equation}
which can be evaluated to obtain
\begin{equation}
C(t)=\langle \phi(t,r)\,{\cal O}_R(t'')\rangle 
=\frac{1}{2}\,\frac{1}{r-\sqrt{r^2-R^2}\,\cosh\left(\frac{t-t''}{R}\right)}\,.
\label{eq:phi_OR_final}
\end{equation}
In a similar manner, one can also find 
\begin{equation}
\langle \phi(t,r)\,{\cal O}_L(t'')\rangle 
=\frac{1}{2}\,\frac{1}{r+\sqrt{r^2-R^2}\,\cosh\left(\frac{t-t''}{R}\right)}\,.
\label{eq:phi_OL_final}
\end{equation}

\subsection{Late time cut-off and failure of microcausality}
\label{subsec:cutoff_modified}

\label{subsubsec:def_cutoff_corr}

\begin{figure}[h]
\centering
\includegraphics[height=0.4\textheight,width=0.75\textwidth]{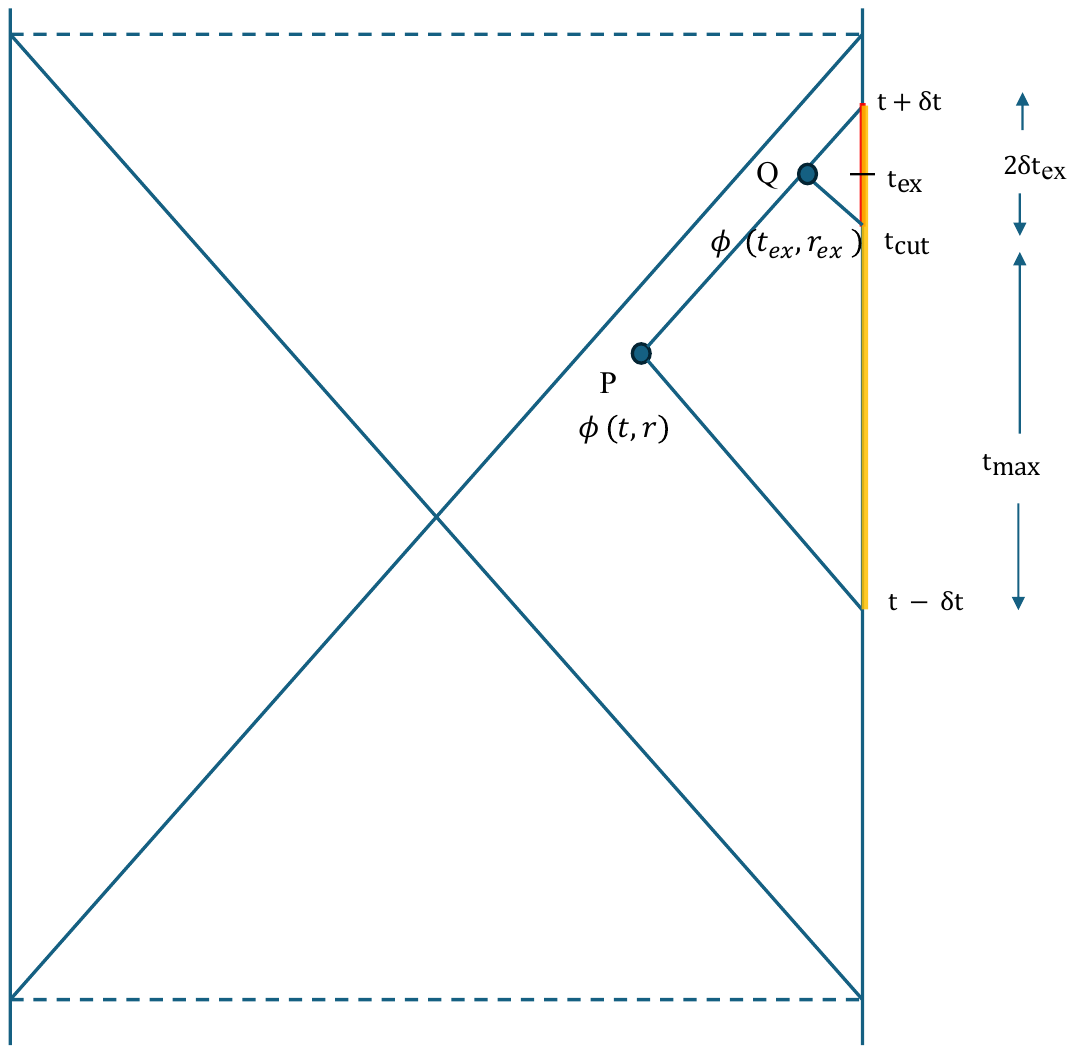}
\caption{Finite $N$ excisions in bulk reconstruction.}
\label{fig:cut-off}
\end{figure}

As mentioned before, at late boundary times the semiclassical boundary to boundary correlator decay by design and gives rise to bulk to boundary correlators \eqref{eq:phi_OR_final} and \eqref{eq:phi_OL_final}. Thus the usual HKLL formalism is not supposed to capture the fluctuating finite $N$ effects, and indeed the bulk reconstruction formalism breaks down if we try to force local HKLL reconstruction at finite $N$. However, one possible way out is to define a non-semiclassical field which is blind to these fluctuations, and carry out their boundary to bulk dictionary in the standard HKLL fashion. 
In order to achieve this, one can impose a cut-off $t_{\rm cut}$ on the upper limit of the smearing integral for bulk fields located close to the horizon. This modifies the bulk fields from their semiclassical counterparts (those which appear in \eqref{eq:phi_OR_final} or \eqref{eq:phi_OL_final} above) to $\phi_{\rm mod}$:
\begin{equation}
\langle \phi_{\rm mod}(t,r)\,{\cal O}_R(t'')\rangle
=\frac{1}{4R^2}\int_{t-\delta t}^{t_{\rm cut}}dt'\,\frac{1}{1-\cosh\left(\frac{t'-t''}{R}\right)}\,.
\label{eq:phi_mod_def}
\end{equation}
This integration region is shaded in green in figure \ref{fig:cut-off}. Here $t_{\rm cut}$ should be understood as an effective endpoint reflecting the breakdown of the semiclassical approximation at very large boundary-time separations. In particular, when $r\sim R$, the integration range is no longer symmetric about $t$, and the resulting bulk--boundary correlator will differ from the semiclassical expression \eqref{eq:phi_OR_final} by a term that depends only on the excised late-time segment. For an arbitrary $t_{\rm cut}<t+\delta t$ we therefore have (e.g. we can take $t_{\rm cut}=t-\delta t+ t_{\rm max}$)

\begin{figure}[h]
\centering
\includegraphics[height=0.4\textheight,width=0.75\textwidth]{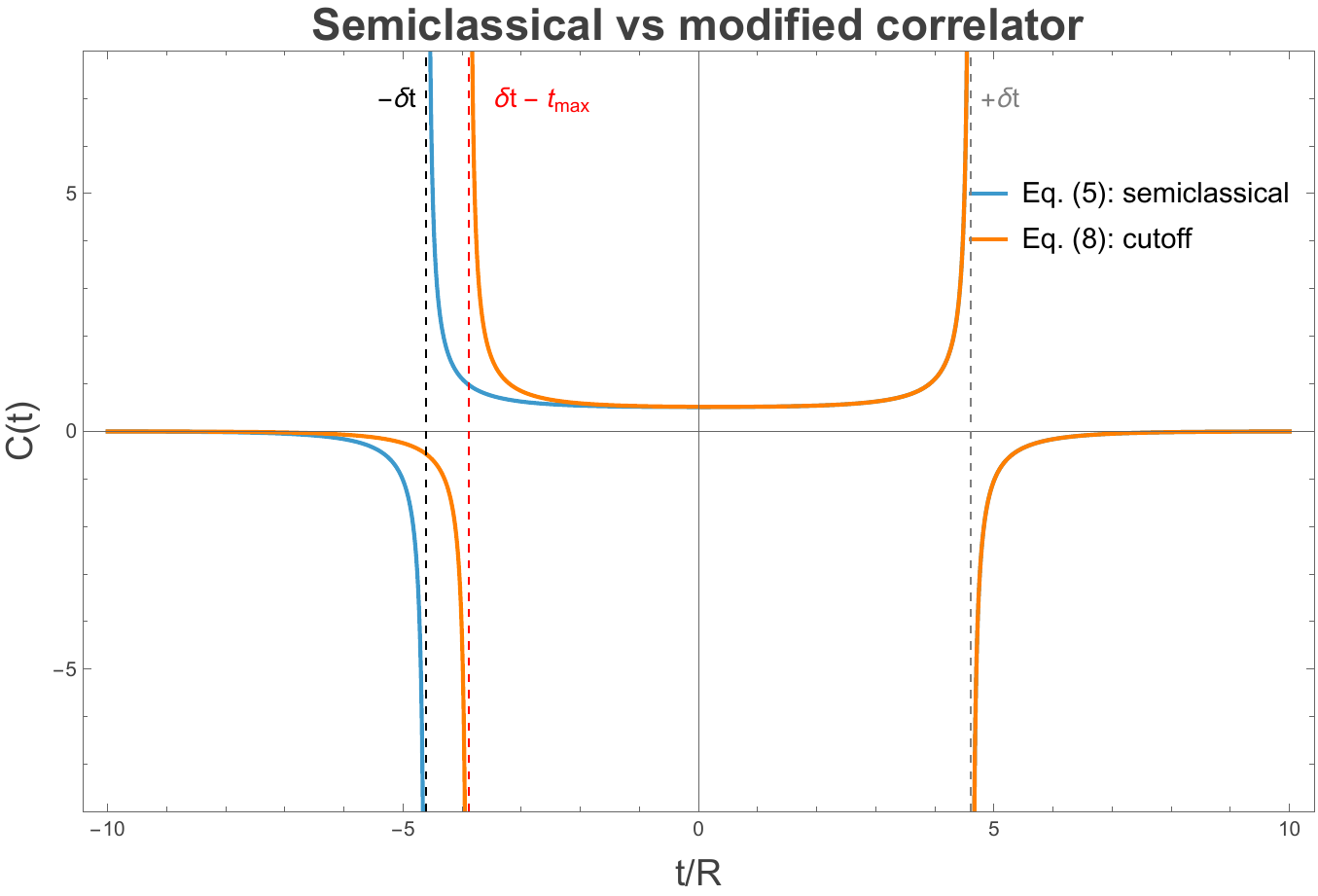}
\caption{Bulk to boundary correlators involving local and non-local bulk fields. The boundary operator is located at $t''=0$. The singularities are the usual lightcone divergences. We have also taken $r=1.0002$, $R=1$ and $t_{\rm max}=8.5$.}
\label{fig:compare_corr}
\end{figure}

\begin{equation}
C(t)=\langle \phi_{\rm mod}(t,r)\,{\cal O}_R(t'')\rangle
=\frac{1}{4R}\,\coth\!\left(\frac{t_{\rm cut}-t''}{2R}\right)
-\frac{1}{4R}\,\coth\!\left(\frac{t-\delta t-t''}{2R}\right)\,.
\label{eq:phi_mod_final}
\end{equation}
The difference between the correlators \eqref{eq:phi_mod_final} and \eqref{eq:phi_OR_final} has been depicted in figure \ref{fig:compare_corr}.

At the operator level such a cut-off now provides a modified bulk field $\phi_{\rm mod}$ given by
\begin{equation}
\phi_{\rm mod}(t,r)=\frac{1}{2}\int_{t-\delta t}^{t_{\rm cut}}dt'\,{\cal O}(t')=\phi (t,r)-\frac{1}{2}\int_{t_{\rm cut}}^{t+\delta t}dt'\,{\cal O}(t')\,.
\label{eq:phi_mod_operator}
\end{equation}
Thus, the modified operator differs from the semiclassical HKLL operator by a subtraction supported entirely on the excised segment of the boundary-time interval.
Alternatively, one can treat the subtracted interval $[t_{\rm cut},t+\delta t]$ as a centered smearing interval by defining (this range corresponds to the orange segment in figure \ref{fig:cut-off})
\begin{equation}
t_{\rm ex}\equiv \frac{t_{\rm cut}+(t+\delta t)}{2}=\frac{t+t_{\rm cut}+\delta t}{2},
\qquad
\delta t_{\rm ex}\equiv \frac{(t+\delta t)-t_{\rm cut}}{2}=\frac{t+\delta t-t_{\rm cut}}{2}\,.
\label{eq:tex_dtex}
\end{equation}
In terms of these variables\footnote{This equality will not be strictly valid for the analysis of the massive scalars, even though the excision process should be implemented identically for all bulk fields. The absence of the HKLL Kernel for massless scalars makes this identification precise.\label{fn:massless}}
\begin{equation}
\phi_{\rm mod}(t,r)= \phi (t,r)-\phi (t_{\rm ex},r_{\rm ex})\qquad\text{with}\qquad r_{\rm ex}\equiv R\coth\left(\frac{\delta t_{\rm ex}}{R}\right)=R\coth\!\left(\frac{t+\delta t-t_{\rm cut}}{2R}\right)\,.
\label{eq:phi_mod_excision}
\end{equation}
Since $t_{\rm ex}$ and $\delta t_{\rm ex}$ are determined by the excised interval endpoints, the corresponding bulk point $(t_{\rm ex},r_{\rm ex})$ generically lies close to the boundary in the asymptotic future region of the Poincar\'e patch \cite{Kabat:2014kfa}. In particular, as shown in \cite{Kabat:2014kfa}, in terms of the two dimensional Poincar\'{e} coordinates 
\begin{equation}
	ds^2=\frac{R^2}{Z^2}(-dT^2+dZ^2)\,,
\end{equation}
the excision region in the boundary Rindler time $\delta t_{\rm ex}$ gets translated to a bulk excised region
\begin{equation}
Z_{\rm ex}\approx 2R\,e^{-t_{\max}/R} \qquad\text{with}\qquad t_{\rm ex}=t+\frac{t_{\max}}{2}\,.
\label{eq:Zex_final}
\end{equation}
In terms of the entropy $S$ (with $\beta=2\pi R$), one may also write the above as 
\begin{equation}
Z_{\rm ex}\approx 2R\,e^{-S/(2\Delta)}\,.
\label{eq:Zex_entropy}
\end{equation}
Near the boundary, the second term of \eqref{eq:phi_mod_excision} scales as 
\begin{equation}
\phi(t_{\rm ex},r_{\rm ex})=\phi(T\approx R,Z_{\rm ex})=(Z_{\rm ex})^{\Delta}\,{\cal O}(T\approx R)\,.
\label{eq:Phi_semi_scaling}
\end{equation}
One therefore has the `non-perturbatively' modified bulk field as
\begin{equation}
\phi_{\rm mod}(r,t)=\Phi_{\rm Rindler}(r,t)-\text{(operator localized near $T=R$ with weight $Z_{\rm ex}^{\Delta}$)}\,.
\label{eq:Phi_mod_summary}
\end{equation}
From this viewpoint, the cut-off modified map is not merely a truncation; it is equivalent (for $\Delta=1$) to subtracting a specific semiclassical bulk field supported near the Poincar\'e boundary at late times. The order of non-locality goes as $\mathcal{O}(e^{-S/2})$, and as discussed in \cite{Kabat:2014kfa}, it provides a heuristic justification of why non-perturbative effects start to set in around the Page time of black hole evaporation.

Given the deviation from semiclassical local bulk fields (as given by \eqref{eq:phi_mod_operator} or \eqref{eq:Phi_mod_summary}), it is clear that the modified fields will fail to be local. Considering a boundary operator ${\cal O}(t_b)$ at a boundary time $t_b$, which is spacelike separated from $\phi_{\rm mod}(r,t)$, we can quantify this failure by computing
\begin{equation}
\bigl[\phi_{\rm mod}(t,r),{\cal O}(t_b)\bigr]
=
\bigl[\phi_{\rm semi}(t,r),{\cal O}(t_b)\bigr]
-\frac{1}{2}\int_{t_{\rm cut}}^{t+\delta t}dt'\,\bigl[{\cal O}(t'),{\cal O}(t_b)\bigr]\,.
\label{eq:commutator_split}
\end{equation}
This gives the amount of non-locality to be 
\begin{equation}
\bigl[\phi_{\rm mod}(t,r),{\cal O}(t_b)\bigr]
=
-\frac{1}{2}\int_{t_{\rm cut}}^{t+\delta t}dt'\,\bigl[{\cal O}(t'),{\cal O}(t_b)\bigr]\,.
\label{eq:commutator_excised}
\end{equation}

In the framework of AdS/CFT duality, it is therefore natural to ask what sort of geometric modifications are implemented due to these boundary late time cut-off procedures. In general, it is a hard question, as we are effectively looking for non-perturbative spacetime saddles. However one aspect that appears to be true is that the geometry gets modifications due to late time cut-off in the boundary time. Following that cut-off to the bulk, this seems to be a near-horizon modification. Motivated by this, in the following, we will provide one possible spacetime model that may quantify such a `non-perturbatively' modified geometry, namely the wormholes of Damour-Solodukhin type \cite{Damour:2007ap} (dubbed as type II wormholes in \cite{Potaux:2023fwm}). They were introduced to precisely circumvent the bulk quasi-normal modes, which can be used to probe the (non-) unitarity of black hole physics. There is a vast literature on this subject starting from the work of Maldacena \cite{Maldacena:2001kr}, although directly relevant to us is e.g.~the treatment in \cite{Solodukhin:2004rv,Solodukhin:2005qy}. The most interesting feature of these wormhole geometries is that the would-be horizon is replaced by a smooth throat, rendering a finite tortoise/optical depth. In fact, in four-dimensional spacetimes, these sort of near-horizon modifications can also be shown to arise analytically from backreactions of conformal anomalies, with the conformal fields in the Boulware state \cite{Berthiere:2017tms}.\footnote{In fact, even in two spacetime dimensions, the choice of the Boulware state almost always yields a horizonless geometry. See the analyses of \cite{Potaux:2021yan,Germani:2015tda,Potaux:2023fwm,delRio:2024fxn,Maranon-Gonzalez:2025rhx} e.g.}  As a result of this throat, if one performs `near-horizon' bulk reconstruction in these geometries, the smearing domain becomes naturally compact without requiring an external $t_{\rm cut}$. As a result, the regulated bulk operator arises from solving the bulk wave equation on a smooth spacetime. In the next section, we'll show this analysis in detail.

\section{Bulk reconstruction in the presence of throats}
\label{sec:AdS2_DSW_cutoff}

In the previous section, we reviewed how the finite $N$ non-localities can be realized within the bulk reconstruction framework by \emph{truncating} the HKLL smearing integral and rewriting the result as a subtraction of an `excised' semiclassical contribution (localized near the Poincar\'e boundary at late times). 
In this section, we will proceed from a bulk view-point. Our main goal will be to carry out a similar process to bulk reconstruction in a geometry which underwent a near-horizon modification, signaling a late time alteration of boundary physics. Intuitively, this connection between the boundary late time excision, and the resulting bulk near-horizon modification can be understood using bulk causality. If the bulk geometry is modified anywhere away from the near-horizon limit, due to bulk causality, its effects would be felt at a boundary time scale much shorter than $t_{\rm max}$.

 There are obviously many possible candidates for such a geometry, but we will in particular choose the wormholes of Damour--Solodukhin (DS) type, which sometimes arise as a non-perturbative solution of exact semiclassical equations, sourced by conformal anomaly \cite{Berthiere:2017tms}.\footnote{Usually these solutions appear quite generically for bulk matter fields satisfying Boulware boundary condition.  Under these conditions, the near boundary or the asymptotic nature of the bulk spacetime remains unchanged. And by definition, they create a large backreaction near the horizon. These geometries also play an interesting role in computing extractable entanglement in field theory \cite{Anegawa:2021osi,Anegawa:2022pce}.} These have the special features of restoring unitarity by producing normal modes, and as we will see, these geometries implement the previously-mentioned boundary finite-support mechanism in a very natural way. As noted in \cite{Solodukhin:2005qy}, these types of geometries also closely produce the brick wall type non-perturbative features (for some relevant comments and references on brick walls, see section \ref{sec:numerics_SFF_3D}).

\subsection{Bulk dynamics in the presence of the throat parameter $\lambda$}
\label{subsec:DSW_geometry_AdS2}

Even though the original work on the DS wormholes featured an asymptotically AdS$_3$ geometry, here we will first tackle the bulk reconstruction in two spacetime dimensions (we will call it DS$_2$ for simplicity). The higher dimensional generalizations lead to analytic complications, which we will deal with in the next section, to some extent.\footnote{In three dimensions, the analogous wormhole background yields a separated radial ODE with four regular singular points, i.e.~a general Heun equation. Local solutions are naturally expressed in terms of $\mathrm{HeunG}$ with parameters fixed by Frobenius analysis around the singular points.} For now, confining our analyses in two dimensions, we start by writing down a dimensionally reduced DS wormhole geometry\footnote{Throughout this paper, DS wormhole refers to a Lorentzian, static, horizonless black-hole foil geometry in which the would-be horizon is replaced by a smooth throat. It should not be confused with a Euclidean replica wormhole.}
\begin{equation}
ds^2=-\frac{r^2-r_h^2+\lambda^2}{R^2}\,dt^2+\frac{R^2}{r^2-r_h^2}\,dr^2\,.
\label{eq:AdS2_DSW_metric_intro}
\end{equation}
Here the parameter $\lambda$ resolves the would-be horizon \(r=r_h\) into a smooth throat for any $\lambda\neq 0$. Introducing the tortoise coordinate
\begin{equation}
r_*=\int \frac{R^2\,dr}{\sqrt{(r^2-r_h^2)(r^2-r_h^2+\lambda^2)}} \,,
\label{eq:AdS2_tortoise_intro}
\end{equation}
one finds that $r_*$ has a \emph{finite} minimum $r_*^{\max}$ (in contrast with the black hole horizon, where $r_*\to-\infty$) as we approach the throat. On the other hand, we can fix the integration constant to make $r^*\to 0$ near boundary. In the regime $\lambda\ll r_h$, (neglecting all higher order \(\lambda^2\) terms), we have
\begin{equation}
r_*^{\max}\approx-\frac{R^2}{r_h}\log\frac{4r_h}{\lambda}\,,
\qquad r_*\in(r_*^{\max},0)\,.
\label{eq:rstarmax_intro}
\end{equation}
Comparing with the previous section (e.g.~see below \eqref{eq:HKLL_AdS2_intro}), we identify this finiteness as the geometric counterpart of a cut-off on large time separations: it bounds how far into the boundary-time domain the reconstruction kernel can reach.

The HKLL consequence is immediate. In this asymptotically AdS wormhole geometry, let us consider a free, local bulk field. In doing so, we are assuming that the bulk reconstruction method, and the usual holographic dictionary go through for small enough $\lambda$. 
For a massless scalar, the scalar Klein-Gordon equation on the DS wormhole becomes a flat spacetime wave equation in $(t,r_*)$ coordinates (for more details see appendix \ref{subsec:rstar_AdS2}, where we have also analyzed the wave equation for the massive scalars) 
\begin{equation}
\left(-\partial_t^2+\partial_{r_*}^2\right)\phi(t,r_*)=0\,,
\label{eq:wave_eq_intro}
\end{equation}
with local normalizable modes behaving as $\phi\stackrel{r^*\to 0}{\sim} e^{-i\omega t}\sin(\omega r_*)$ near the boundary \cite{Sarkar:2014jia}. Via the extrapolate dictionary, this leads to a smearing representation
\begin{equation}
\phi(t,r_*)=\frac{1}{2}\int_{t-r_*}^{t+r_*}dt'\,{\cal O}(t')\,.
\label{eq:HKLL_wormhole_intro}
\end{equation}
Since $r_*$ is always finite, the smearing support is \emph{inherently finite}, and as the bulk point approaches the throat the maximum time separation is controlled by $r_*^{\max}$. In particular, as $r_*\to r_*^{\max}$, the smearing window approaches $[t-r_*^{\max},\,t+r_*^{\max}]$. So the geometry imposes an effective large-separation cut-off
\begin{equation}
|t'-t|\;\lesssim\;|r_*^{\max}|\;\sim\;\frac{R^2}{r_h}\log\!\frac{4r_h}{\lambda}\,,
\label{eq:effective_cutoff_intro}
\end{equation}
set dynamically by $\lambda$ rather than imposed by hand. This is the sense in which DS wormholes geometrize finite $N$ motivated cut-off/excision modifications of HKLL \cite{Kabat:2014kfa}, with $r_*^{\max}$ playing the analogous role of $\delta t$.\footnote{Note the difference between \eqref{eq:HKLL_wormhole_intro} and \eqref{eq:phi_mod_operator}. They are strictly not the same formulas for a modified bulk field, as the wormhole modification gives a symmetrical time excision. However, the order of non-locality remains the same between the two approaches, which we show explicitly in the next subsection.} It is important to note that whereas they are local on the wormhole background by construction, from the perspective of the BTZ background \cite{Banados:1992wn}, these fields are automatically non-local. The amount of non-locality will also be same as the cut-off procedure, upon proper identification of the excision parameters. We turn to this identification next.

\subsection{Matching to the finite $N$ cut-off scale and estimating $\lambda$}
\label{subsec:match_lambda}

Equating \eqref{eq:Zex_final} and \eqref{eq:Zex_entropy}, or alternatively using $t_{\max}=\frac{\beta S}{4\pi \Delta}$, we can write 
\begin{equation}
\frac{t_{\max}}{R}=\frac{S}{2\Delta}\,.
\label{eq:tmax_entropy_match}
\end{equation}
Geometrically, as the maximal temporal reach of the wormhole smearing is controlled by $|r_*^{\max}|$, a minimal matching criterion is therefore\footnote{We have consistently used a slightly different estimate for $t_{\max}$ in both this section and the previous one, as compared to \cite{Kabat:2014kfa}. We find it better suited, and it doesn't change the order of non-locality.} 
\begin{equation}
t_{\max}\approx 2\,\delta t = 2\,|r_*^{\max}|=-2\,r_*^{\max}\,.
\label{eq:tmax_match_rstarmax}
\end{equation}
In other words, the excision time scale of the cut-off is identified with the maximal tortoise depth of the wormhole. Using \eqref{eq:rstarmax_intro} we therefore have
\begin{equation}
t_{\max}\approx \frac{2R^2}{r_h}\log\left(\frac{4r_h}{\lambda}\right)
\quad\Rightarrow\quad
\lambda \approx 4r_h\,\exp\!\left(-\frac{r_h}{2R^2}t_{\max}\right)\,.
\label{eq:solve_lambda_log}
\end{equation}
With the usual normalization choice $r_h\simeq R$, this becomes
\begin{equation}\label{eq:lambdaestimate}
	\lambda\approx 4R\,e^{-t_{\max}/2R}\approx e^{-\frac{S}{4\Delta}}\sim e^{-S/4} \qquad \text{which implies}\qquad \lambda^2\sim e^{-S/2}\qquad\text{(for $\Delta=1$)}\,.
\end{equation}
	This is precisely the order of estimate of $\lambda$ made in \cite{Solodukhin:2005qy}, which was later verified in the particular backreaction problem of \cite{Berthiere:2017tms}. Therefore, with these identifications, one now sees that the bulk field that are naturally local in the DS$_2$ wormhole geometry will be a non-local field in AdS$_2$ Rindler (or black hole geometry) by an amount given by $\lambda^2\sim e^{-S/2}$. In the next section, we will again recover an $\mathcal{O}(\lambda^2)$ deviations from BTZ semi-classicality, which is somewhat guaranteed as the wormhole metric ansatz has an $\mathcal{O}(\lambda^2)$ correction on top of the classical geometry. However, as we will see, the precise relation between $\lambda$ and $S$ will depend on the field contents, their conformal dimensions etc.

\section{Story in three dimensions}
\label{sec:3Dheun}

In the previous two sections~\ref{sec:AdS2_cutoff_excision} and \ref{sec:AdS2_DSW_cutoff} we made our central mechanism transparent with the example of asymptotically AdS$_2$ wormholes. We pointed out that a finite $N$ motivated cut-off/excision process can be implemented \emph{geometrically} by a DS$_2$ wormhole throat, without needing to impose an external endpoint $t_{\rm cut}$ in the HKLL integrals. The same geometric ideas persist in three dimensions as well, but now the separated wave equation becomes structurally richer. The radial part is generically non-integrable and reduces to a general Heun equation, which we'll now briefly mention. Upon deriving the exact 3D radial ODE and presenting its reduction to the canonical Heun form, we'll first carry out the bulk reconstruction and will later analyze the near-throat regime to extract the local behavior of modes. 

\subsection{Exact bulk dynamics in DS$_3$ wormholes}
\label{subsec:3D_setup}

The three-dimensional asymptotically AdS DS wormhole metric takes the form
\begin{equation}
ds^2
=
-\frac{r^2-r_h^2+\lambda^2}{R^2}\,dt^2
+\frac{R^2}{r^2-r_h^2}\,dr^2
+r^2\,d\varphi^2\,,
\label{eq:AdS3_wh_metric}
\end{equation}
where \(R\) is the AdS radius, \(r_h\) sets the BTZ scale (with inverse temperature $\beta_H=\frac{2\pi R^2}{r_h}$), and \(\lambda\) controls the wormhole throat deformation. As in the two-dimensional case, the would-be horizon at \(r=r_h\) is resolved into a smooth throat for any \(\lambda\neq 0\). We study a scalar field \(\Phi\) of mass \(m\) obeying
\begin{equation}
(\Box - m^2)\Phi=0\,,
\label{eq:KG3D}
\end{equation}
with the mode ansatz
\begin{equation}
\Phi_{\omega k}(t,r,\phi)=e^{-i\omega t}\,e^{ik\varphi}\,\mathcal{R}_{\omega k}(r)\,.
\label{eq:mode_ansatz_3D}
\end{equation}
Separating the variables, the dynamics reduces to a one-dimensional spectral problem for \(\mathcal{R}_{\omega k}(r)\) with frequency \(\omega\) and angular momentum \(k\). Unlike in two-dimensions, where the massless equation is conformally trivial in \((t,r_*)\), the 3D radial equation retains nontrivial singular-point structure, and it is this structure that ultimately produces the Heun form. The equation for the radial part is given by 
\begin{equation}
\frac{1}{r}\sqrt{\frac{r^2-r_h^2}{r^2-r_h^2+\lambda^2}}\;
\frac{d}{dr}\!\left(
r\sqrt{(r^2-r_h^2)(r^2-r_h^2+\lambda^2)}\;\frac{d\mathcal{R}_{\omega k}}{dr}
\right)
+
R^2\left(
\frac{R^2\omega^2}{r^2-r_h^2+\lambda^2}
-\frac{k^2}{r^2}
-m^2
\right)\mathcal{R}_{\omega k}
=0\,.
\label{eq:radial_r_final_3D}
\end{equation}
Defining some new variables (same as in \eqref{eq:z_zs_def_massive})
\begin{equation}
z = \frac{r_h^2}{r^2}
\qquad \text{and}\qquad
z_s = \frac{r_h^2}{r_h^2-\lambda^2}
\qquad (z_s>1)\,,
\label{eq:z_and_zs_def_3D}
\end{equation}
we can convert \eqref{eq:radial_r_final_3D} to the following form 
\begin{equation}
\mathcal{R}_{\omega k}''(z)
+
\left(
\frac{1}{2(z-1)}+\frac{1}{2(z-z_s)}
\right)\mathcal{R}_{\omega k}'(z)
+
\mathcal{Q}(z)\,\mathcal{R}_{\omega k}(z)=0\,,
\label{eq:Rz_master_3D}
\end{equation}
with
\begin{equation}
\mathcal{Q}(z)
=
\frac{R^2 m^2}{4\,z^2(z-1)}
+
\frac{R^2 k^2}{4 r_h^2\,z(z-1)}
+
\frac{R^4\omega^2}{4(r_h^2-\lambda^2)\,z(z-1)(z-z_s)}\,.
\label{eq:Qz_3D}
\end{equation}
One can try and solve this equation analytically with the near-boundary (\(z=0\)) ansatz $y_{\omega k}(z)=z^{-\alpha_0}\,\mathcal{R}_{\omega k}(z)$, which also makes the connection with the bulk  reconstruction explicit. With this ansatz, near \(z=0\) the indicial equation gives 
\begin{equation}
\alpha_0(\alpha_0-1)-\frac{R^2 m^2}{4}=0
\quad\Longrightarrow\quad
\alpha_0=\frac{1}{2}\left(1\pm \sqrt{1+R^2 m^2}\right)=\alpha_{\pm}\,.
\label{eq:alpha0_3D}
\end{equation}
As usual (and also encountered for the massive scalar case in appendix \ref{subsec:rstar_AdS2}), the two roots correspond to the two independent near-boundary behaviors.
The function $y(z)$ satisfies the canonical general Heun equation
\begin{equation}
y_{\omega k}''(z)
+\left(
\frac{\gamma}{z}+\frac{\delta}{z-1}+\frac{\epsilon}{z-a}
\right)y_{\omega k}'(z)
+\frac{\alpha\beta\, z - q}{z(z-1)(z-a)}\,y_{\omega k}(z)=0\,,
\label{eq:Heun_canonical_3D}
\end{equation}
with the explicit parameters
\begin{align}
a=z_s=\frac{r_h^2}{r_h^2-\lambda^2}\,,\quad
\gamma=2\alpha_0\,,\quad
\delta=\frac{1}{2}\,,\quad
\epsilon=\frac{1}{2}\,,\quad
\alpha=\alpha_0+\frac{iRk}{2r_h}\,,\quad \nonumber \\
\beta=\alpha_0-\frac{iRk}{2r_h}\,,\quad
q=
z_s\left(
\frac{R^2 m^2}{4}
+\frac{R^2 k^2}{4r_h^2}
+\frac{\alpha_0}{2}
\right)
+\frac{\alpha_0}{2}
-\frac{R^4\omega^2}{4(r_h^2-\lambda^2)}\,.
\label{eq:Heun_params_boxed_3D}
\end{align}
Accordingly, the `normalizable' solution about \(z=0\) can be written as\footnote{All the parameters $q,\alpha,\beta,\gamma$ appearing for the normalizable mode should be understood as the ones defined in \eqref{eq:Heun_params_boxed_3D}, but with the $\alpha_0=\alpha_+$ root.}
\begin{equation}
\mathcal{R}_{\omega k}(z)=z^{\alpha_+}\,
\mathrm{HeunG}\!\left(a=z_s,\; q,\; \alpha,\; \beta,\; \gamma,\; \delta;\; z\right)\,,
\label{eq:R_HeunG_solution_3D}
\end{equation}
together with the `non-normalizable' independent Heun solution corresponding to the second local exponent at \(z=0\). In later sections, the explicit Heun reduction will be used as a precise characterization of the radial problem in terms of a Schr\"odinger/WKB formulation. This will help us obtain an approximate spectra to perform the spectral diagnostics.

In addition, one can also perform a near-throat analysis of the bulk dynamics by passing to the variable 
\begin{equation}
\rho \equiv \sqrt{r^2-r_h^2}
\qquad (\rho\ge 0)\,,
\label{eq:rho_def}
\end{equation}
in terms of which \eqref{eq:radial_r_final_3D} takes the form
\begin{equation}
\frac{1}{R^2}\frac{1}{\sqrt{\rho^2+\lambda^2}}
\frac{d}{d\rho}\!\left(
(r_h^2+\rho^2)\sqrt{\rho^2+\lambda^2}\,\frac{d\mathcal{R}_{\omega k}}{d\rho}
\right)
+
\left(
\frac{R^2\omega^2}{\rho^2+\lambda^2}
-\frac{k^2}{r_h^2+\rho^2}
-m^2
\right)\mathcal{R}_{\omega k}
=0\,.
\label{eq:exact_rho_equation}
\end{equation}
In these variables, one may also analyze the bulk dynamics analytically in the regions $\rho\ll\lambda$ or in the region $\lambda\ll\rho\ll r_h$. We have provided some details of the corresponding solutions in appendix \ref{subsec:near_throat_3D}. 

Introducing the conformal dimension (which is $\gamma=2\alpha_0$ as defined in \eqref{eq:Heun_params_boxed_3D}, but with the $\alpha_0=\alpha_+$ root)
\begin{equation}
\Delta=1+\sqrt{1+R^2m^2}\,,
\label{eq:Delta_def_recon}
\end{equation}
one has
\begin{equation}
\alpha_+=\frac{\Delta}{2},
\qquad
\alpha_-=\frac{2-\Delta}{2}\,.
\label{eq:alpha_pm_Delta_recon}
\end{equation}
Therefore the two near-boundary modes are (as $z\to 0$)
\begin{equation}
\mathcal R_{\omega k}(z)
\sim
A_{\omega k}\,z^{(2-\Delta)/2}
+
B_{\omega k}\,z^{\Delta/2}\,.
\label{eq:boundary_branches_recon}
\end{equation}
The first branch is `non-normalizable'. Here $A_{\omega k}$ and $B_{\omega k}$ comes from the near boundary expansion of the Heun solution \eqref{eq:R_HeunG_solution_3D}.
In particular, the `normalizable' radial mode is given by
\begin{equation}
\mathcal R^{\rm norm}_{\omega k}(z)
=
z^{\Delta/2}\,
\text{HeunG}\!\left(
z_s,\,
q_+,\,
\frac{\Delta}{2}+\frac{iRk}{2r_h},\,
\frac{\Delta}{2}-\frac{iRk}{2r_h},\,
\Delta,\,
\frac12;\,
z
\right),
\label{eq:Heun_normalizable_mode_recon}
\end{equation}
where 
\begin{equation}
q_+
=q|_{\alpha_0=\alpha_+}=
z_s\!\left(
\frac{R^2m^2}{4}
+\frac{R^2k^2}{4r_h^2}
+\frac{\Delta}{4}
\right)
+\frac{\Delta}{4}
-\frac{R^4\omega^2}{4(r_h^2-\lambda^2)}\,.
\label{eq:q_plus_recon}
\end{equation}
Expanding around \(z=0\), one uses
\begin{equation}
\text{HeunG}(a,q,\alpha,\beta,\gamma,\delta;z)
=
1+\frac{q}{a\gamma}\,z+O(z^2)\,,
\label{eq:Heun_series_zero_recon}
\end{equation}
so that
\begin{equation}
\mathcal R^{\rm norm}_{\omega k}(z)|_{z\to 0}
=
z^{\Delta/2}
\left[
1+\frac{q_+}{z_s\Delta}\,z+O(z^2)
\right]\,.
\label{eq:normalizable_mode_series_recon}
\end{equation}
The bulk field on the right patch is then formally expanded as
\begin{equation}
\Phi_R(t,z,\varphi)
=
\sum_{k\in\mathbb Z}\int\frac{d\omega}{2\pi}\,
e^{-i\omega t}e^{ik\varphi}\,
\mathcal R^{\rm norm}_{\omega k}(z)\,
a_{\omega k}
+\text{h.c.}
\label{eq:bulk_mode_sum_right_patch_recon}
\end{equation}
Identifying the boundary operator through the extrapolate dictionary near $z\to 0$
\begin{equation}
\mathcal O(t,\varphi)
=
\lim_{z\to 0}z^{-\Delta/2}\,\Phi_R(t,z,\varphi)\,,
\label{eq:extrapolate_dict_recon}
\end{equation}
and using \eqref{eq:normalizable_mode_series_recon}, we have
\begin{equation}
\mathcal O(t,\varphi)
=
\sum_{k\in\mathbb Z}\int\frac{d\omega}{2\pi}\,
e^{-i\omega t}e^{ik\varphi}\,
a_{\omega k}
+\text{h.c.}
\label{eq:boundary_mode_expansion_recon}
\end{equation}
Identifying $a_{\omega k}$ from the above equation, and substituting it back into \eqref{eq:bulk_mode_sum_right_patch_recon}, one obtains the following relation between the bulk and the boundary operators.
Using the boundary mode expansion, we define 
\begin{equation}
    \mathcal{O}_{\omega k}
    =
    \frac{1}{2\pi}
    \int_{-\infty}^{\infty} dt'
    \int_{0}^{2\pi} d\varphi'\,
    e^{i\omega t'-ik\varphi'}
    \mathcal{O}(t',\varphi')\,.
    \label{eq:boundary_fourier_modes}
\end{equation}
Substituting this expression into the bulk mode expansion, the right-patch bulk field can
be written as
\begin{equation}
    \Phi_{R}(t,z,\varphi)
    =
    \int_{-\infty}^{\infty} dt'
    \int_{0}^{2\pi} d\varphi'\,
    K_{\mathrm{DS}}
    \bigl(t-t',\varphi-\varphi';z\bigr)
    \mathcal{O}(t',\varphi')
    +\mathrm{h.c.}\,,
    \label{eq:bulk_reconstruction_Heun_right_patch}
\end{equation}
where the DS$_3$ smearing kernel is defined by
\begin{align}
    K_{\mathrm{DS}}
    \bigl(t-t',\varphi-\varphi';z\bigr)
    =
    \frac{z^{\Delta/2}}{2\pi}
    \sum_{k\in\mathbb{Z}}
    \int_{-\infty}^{\infty}
    \frac{d\omega}{2\pi}\,
    e^{-i\omega(t-t')+ik(\varphi-\varphi')}
    \,\text{HeunG}\left(
        z_s,
        q_{+},
        \frac{\Delta}{2}+\frac{iRk}{2r_h},
        \frac{\Delta}{2}-\frac{iRk}{2r_h},
        \Delta,
        \frac{1}{2};
        z
    \right).
    \label{eq:DS_smearing_kernel}
\end{align}
This is the formal right-patch reconstruction in terms of the normalizable Heun mode. If one further imposes the smooth-throat condition at \(z=1\), the allowed values of \(\omega\) become discrete. 

In order to make an estimate of the deviation of the resulting bulk field from the semiclassical BTZ field, we now expand the smearing kernel \(K_{\rm DS} \) for the DS$_3$ geometry locally for small throat parameter \(\lambda\). In this case, the singular point \(z=z_s\) approaches \(z=1\) in the BTZ limit (see e.g. \eqref{eq:Qz_3D}). 
This behavior can be reproduced by series expanding $K_{\rm DS}$ in the parameter \(\varepsilon=\lambda^2/r_h^2\). Doing so, one finds
\begin{equation}
K_{\rm DS}(z;\lambda)
=
K_{\rm BTZ}(z)
+
\varepsilon\,K_{\rm corr}(z)
+
O(\varepsilon^2)\,,
\label{eq:DS_kernel_small_lambda_split}
\end{equation}
where the leading piece is precisely the BTZ hypergeometric kernel
\begin{equation}
K_{\rm BTZ}(z)
=\frac{1}{2\pi}
    \sum_{k\in\mathbb{Z}}
    \int_{-\infty}^{\infty}
    \frac{d\omega}{2\pi}\,
    e^{-i\omega(t-t')+ik(\varphi-\varphi')}
z^{\Delta/2}
(1-z)^s
{}_2F_1
\left(
\frac{\Delta}{2}+s+\frac{iRk}{2r_h},
\frac{\Delta}{2}+s-\frac{iRk}{2r_h};
\Delta;z
\right)\,,
\label{eq:BTZ_kernel_from_DS_limit}
\end{equation}
with \( s=\pm\frac{iR^2\omega}{2r_h}\,\).
The first order correction is therefore explicitly \(\lambda^2\) suppressed and is generated by the displacement of the Heun singularity together with the induced shift of the accessory parameter $q$ (for $q=q_+$)
\begin{align}
K_{\rm corr}(z)
=\frac{1}{2\pi}
    \sum_{k\in\mathbb{Z}}
    \int_{-\infty}^{\infty}
    \frac{d\omega}{2\pi}\,
    e^{-i\omega(t-t')+ik(\varphi-\varphi')}
z^{\Delta/2}
\left[
\partial_a
+
\left(
q_{\rm BTZ}
-
\frac{\Delta}{4}
\right)
\partial_q
\right]\nonumber\\
\times \text{HeunG}
\left(
a,q,
\frac{\Delta}{2}+\frac{iRk}{2r_h},
\frac{\Delta}{2}-\frac{iRk}{2r_h},
\Delta,\frac12;z
\right)
\bigg|_{a=1,q=q_{\rm BTZ}}\,,
\label{eq:DS_kernel_correction}
\end{align}
with
\begin{equation}
q_{\rm BTZ}
=
\frac{\Delta^2}{4}
+
\frac{R^2k^2}{4r_h^2}
-
\frac{R^4\omega^2}{4r_h^2}\,.
\label{eq:qBTZ_small_lambda_split}
\end{equation}
Thus, mode by mode, the DS reconstruction naturally takes the form of the BTZ reconstruction plus a finite $\mathcal{O}(\lambda^2)$ deformation
\begin{equation}
\Phi_R^{\rm DS}
=
\Phi_R^{\rm BTZ}
+
\frac{\lambda^2}{r_h^2}\delta\Phi_R
+
O\left(\frac{\lambda^4}{r_h^4}\right)\,.
\label{eq:DS_field_BTZ_plus_correction}
\end{equation}
So, just like in the two-dimensional case, this above relation again provides an exponentially suppressed deviation from semiclassical, local bulk field in BTZ. However, as mentioned earlier, the precise matching between $\lambda$ and the entropy $S$ depends crucially on the details of the theory involved. For example, for a massless scalar ($\Delta=d$), a computation similar to section \ref{subsec:match_lambda} suggests the order of deviation to be $\lambda^2\sim e^{-S/4}$. We have provided a detailed derivation of equations \eqref{eq:DS_kernel_small_lambda_split} through \eqref{eq:DS_field_BTZ_plus_correction} in appendix~\ref{subsec:non_locality_est}. We have also separately studied the near-throat bulk dynamics in appendix~\ref{subsec:near_throat_3D}. Note that in order to derive the non-locality (failure of micro-causality), we have to compute $\frac{\lambda^2}{r_h^2}\left[\delta\Phi_R,\mathcal{O}(t_b,\varphi)\right]$, where $\mathcal{O}(t_b,\varphi)$ is a spacelike separated (from $\Phi_R^{\rm DS}$) boundary operator. Even though there is no reason for this commutator to be zero generically (just like the commutator on the RHS of \eqref{eq:commutator_excised} is non-zero generically), we haven't performed this computation due to complications associated to the Heun function appearing in \eqref{eq:DS_kernel_correction}.

\subsection{Schr\"odinger reduction and effective potential}
\label{sec:WKB_SFF_3D}

In the previous subsection we found that just like in the two-dimensional case, a deviation from semiclassicality in the bulk reconstruction in the BTZ background may alternatively be viewed as originating from modifying the near-horizon geometry. This modification, due to bulk causality, affects all the way to the late boundary time. From this perspective it almost proxies as a stretched horizon. Some recent works \cite{Das:2022wcj,Das:2023btz,Das:2023simp,Das:2023frm,Krishnan:2023nms} performed a similar probe analysis in the background of brick walls in order to find the resulting spectral form factor. Their results curiously yielded a dip-ramp-plateau structure, that are usually obtained in the studies of random matrix theory, and more recently in the studies of gravitational path integral \cite{Cotler:2016fpe,Maldacena:2016upp,Saad:2019lba,Stanford:2019vob}. Essentially, one rewrites the bulk equation in the Schr\"odinger form to extract the effective potential. If the effective potential supports bound states, one implements the Bohr-Sommerfeld quantization to obtain the discrete frequency modes. For the black hole mimickers a similar approach was also taken in \cite{Solodukhin:2005qy}, although the analysis there was carried out under certain approximations. From this section onwards, we improve on that analysis, which will ultimately help us obtain the exact spectral form factor behavior in these near-horizon modified backgrounds.

\begin{figure}
\centering
\includegraphics[width=0.75\textwidth]{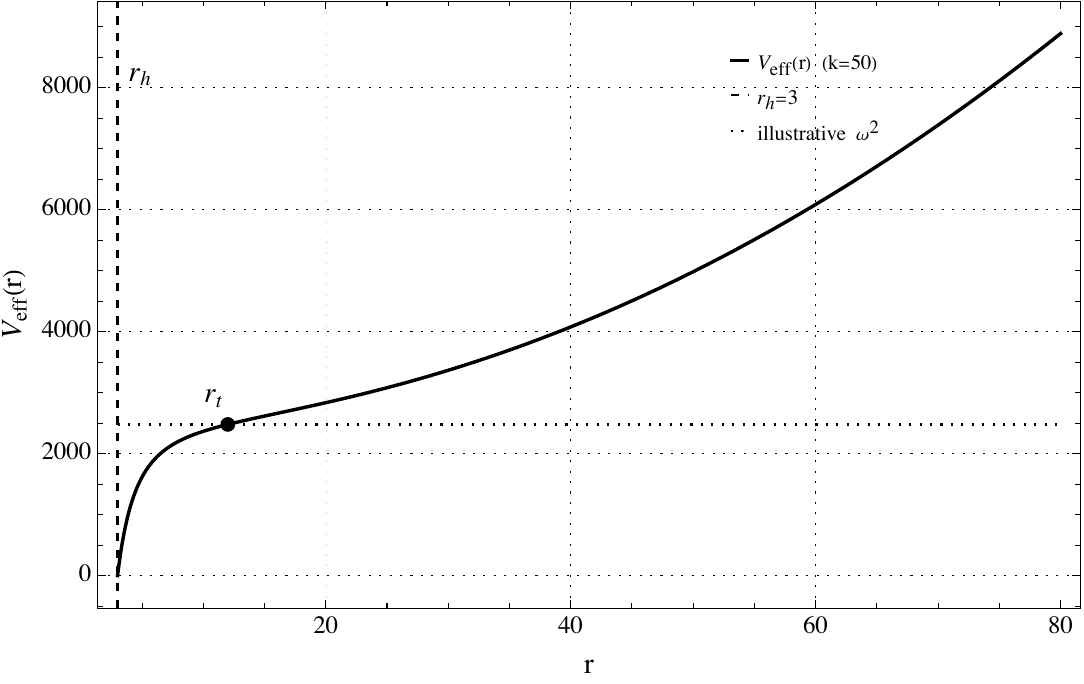}
\caption{One-sided effective potential \(V_{\rm eff}(r)\) for representative values \(\omega^2 \sim 2475\) with \(k=50\). The vertical dashed line marks the throat \(r=r_h\). Here we have also taken $\lambda=5\times 10^{-4}$, $r_h=3$ and $R=1$.}
\label{fig:Veff_r}
\end{figure}

Working with the wormhole metric \eqref{eq:AdS3_wh_metric}, we start by defining
\begin{equation}
\alpha(r)\equiv \sqrt{r^2-r_h^2+\lambda^2}\,,
\qquad
\beta(r)\equiv \sqrt{r^2-r_h^2}\,,
\label{eq:alpha_beta_def_rewrite}
\end{equation}
through which the tortoise coordinate \(r_*\) is defined as
\begin{equation}
\frac{dr_*}{dr}=\frac{R^2}{\alpha(r)\beta(r)}\,.
\label{eq:tortoise_def_rewrite}
\end{equation}
After separation of variables \(\Phi_{\omega k}=e^{-i\omega t}e^{ik\varphi}\psi(r)\), the massless radial equation takes the form
\begin{equation}
\overset{**}\psi+\frac{\overset{*}r}{r}\,\overset{*}\psi+\left(\omega^2-\frac{k^2\alpha^2}{R^2 r^2}\right)\psi=0\,.
\label{eq:eq3_app}
\end{equation}
Here $\overset{*}r\equiv dr/dr_*$, \(\overset{*}\psi \equiv d\psi/dr_*\) and \(\overset{**}\psi\equiv d^2\psi/dr_*^2\). One removes the first-derivative term by the standard redefinition
\begin{equation}
\psi(r_*)=\frac{\phi(r_*)}{\sqrt{r}}\,,
\label{eq:psi_phi_def_rewrite}
\end{equation}
in terms of which the radial KG equation takes the Schr\"odinger form
\begin{equation}
\frac{d^2\phi}{dr_*^2}+\bigl(\omega^2-V_{\rm eff}(r)\bigr)\phi=0\,.
\label{eq:schroedinger_form_rewrite}
\end{equation}
Here \(r\) is understood as a function of \(r_*\) via \eqref{eq:tortoise_def_rewrite}.
We can therefore read off the Schr\"odinger potential as \(V(r)\equiv V_{\rm eff}(r)-\omega^2\) with (below $\overset{**}r=\frac{d^2r}{dr{_*}^2}$)
\begin{align}\label{eq:Veff_full_rewrite}
	V_{\rm eff}(r)&=\frac{k^2\alpha^2}{R^2 r^2}+\frac{\overset{**}r}{2r}-\frac{1}{4}\left(\frac{\overset{*}r}{r}\right)^2\nonumber\\
	&=\frac{k^2\bigl(r^2-r_h^2+\lambda^2\bigr)}{R^2 r^2}+\frac{2(r^2-r_h^2)+\lambda^2}{2R^4}-\frac{(r^2-r_h^2)\bigl(r^2-r_h^2+\lambda^2\bigr)}{4R^4 r^2}\,.
\end{align}
The behavior of $V_{\rm eff}$ with $r$ has been given in figure \ref{fig:Veff_r}. 

\subsection{Effective potential and Bohr--Sommerfeld quantization}
\label{subsec:WKB_quantization_rewrite}

\begin{figure}
\centering
\begin{subfigure}[t]{0.49\textwidth}
  \centering
  \includegraphics[width=\textwidth]{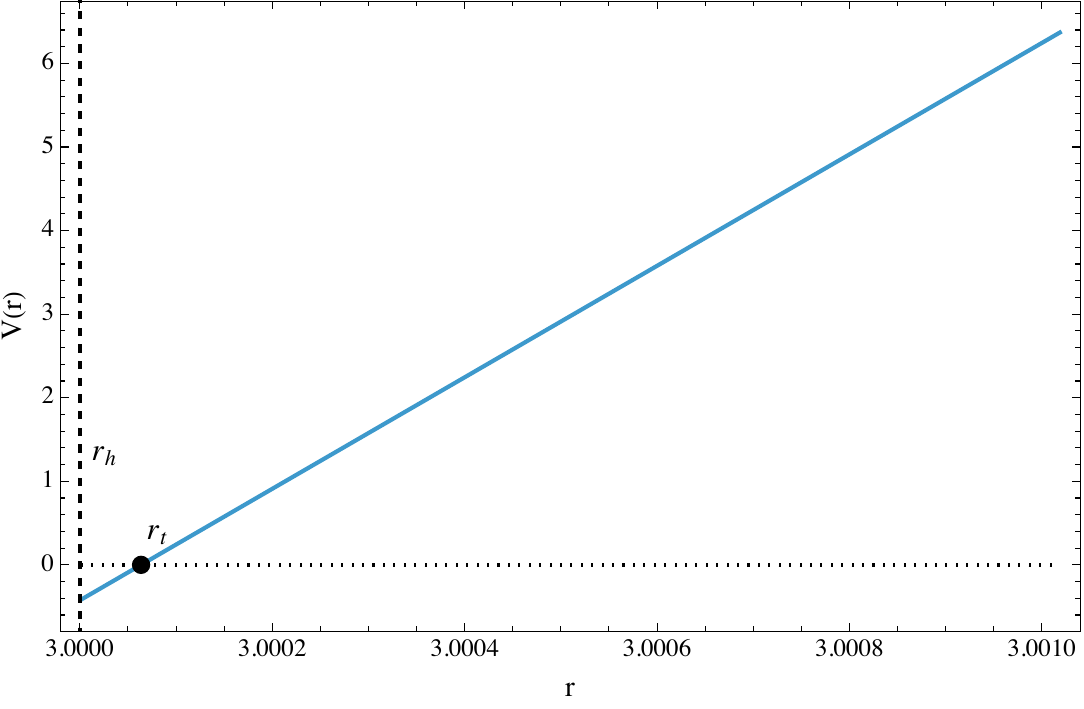}
  \caption{\(V(r)\) vs.\ \(r\).}
  \label{fig:Veff_r_sidepanel}
\end{subfigure}\hfill
\begin{subfigure}[t]{0.49\textwidth}
  \centering
  \includegraphics[width=\textwidth]{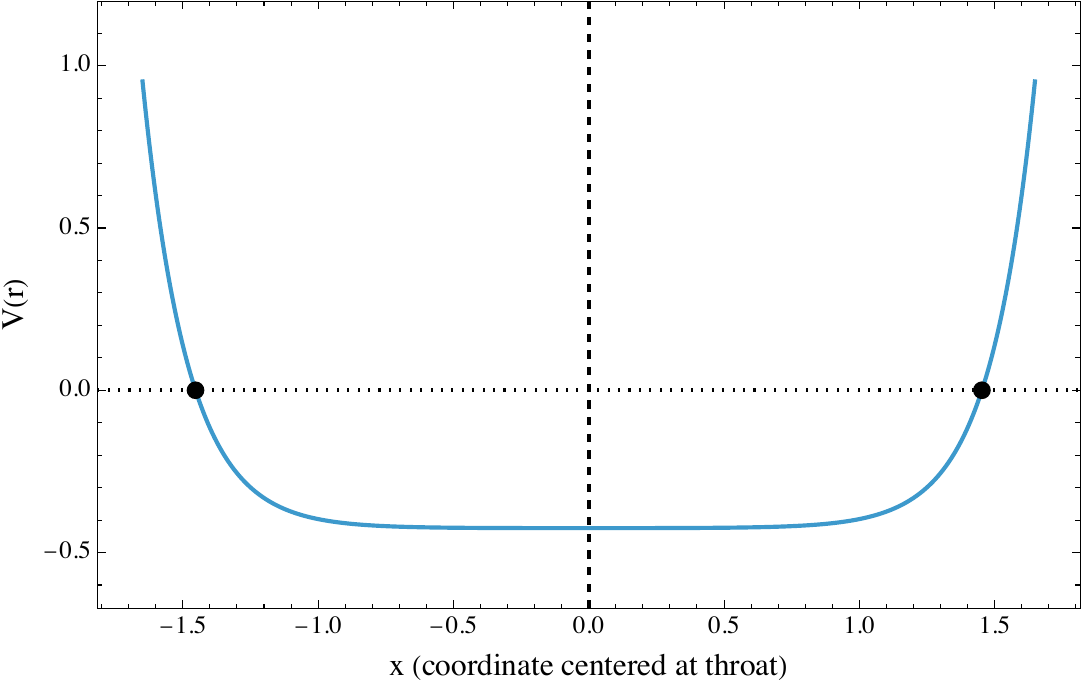}
  \caption{\(V(x)\) vs.\ \(x\).}
  \label{fig:Veff_x_sidepanel}
\end{subfigure}
\caption{Schr\"odinger potential in the radial coordinate and in tortoise coordinate (\(x\equiv r_\ast\)). The tortoise-coordinate representation is symmetric under \(x\to -x\), so turning points occur at \(x=\pm x_t\). We have taken $R=1$, $\omega=0.652$, $k=100$, $\lambda=5\times 10^{-4}$ and $r_h=3$.}
\label{fig:Veff_rstar}
\end{figure}

Before turning to the spectral analysis, it is useful to visualize the effective potential \(V_{\rm eff}(r)\) in \eqref{eq:Veff_full_rewrite}, since its qualitative structure controls turning points, WKB phase space, and hence the density and organization of levels.  As is clear from \eqref{eq:Veff_full_rewrite}, $V_{\rm eff}(r)$ is an even function of $\rho$ as defined in \eqref{eq:rho_def}. Motivated by this, in the following, we'll use a parametric coordinate \(x\) that spans both sides of the wormhole centered at the location of the throat (i.e.~$x=0$ at \(r=r_h\)). Its definition is essentially equivalent to that of \(r_*\) (as given by \eqref{eq:AdS2_tortoise_intro}) written as a function of $\rho$.
\begin{equation}
\frac{dx}{d\rho}
=
\frac{R^2}{\sqrt{r_h^2+\rho^2}\,\sqrt{\rho^2+\lambda^2}}\,.
\end{equation}
More precisely, in our numerical implementations we'll define the throat-centered coordinate \(x\) by
\begin{equation}
x(\rho)\equiv \int_{0}^{\rho}\frac{R^2\,ds}{\sqrt{r_h^2+s^2}\,\sqrt{s^2+\lambda^2}}\,,
\label{eq:xu_def}
\end{equation}
with the boundary condition \(x(0)=0\) for \(r=r_h\). Equivalently, \(x\) can be understood as the \emph{shifted} tortoise coordinate in which the throat is at the origin.

For our expression \eqref{eq:Veff_full_rewrite}, figure \ref{fig:Veff_r} provides a visualization of the allowed region which lies between \(r=r_h\) to \(r=r_t\). In reality, the region is very narrow, as expected from the estimation of \(\lambda\), and has been shown for some given values in figure \ref{fig:Veff_r_sidepanel}. In terms of the centered coordinate \(x\), one sees the symmetric shallow well in figure \ref{fig:Veff_x_sidepanel}. 

Given the symmetric shallow well in the $x$ coordinate, we are now ready to perform the Bohr-Sommerfeld quantization. For fixed \((k,\lambda)\), we see that the effective potential has two turning points. Back in $r$ coordinate, we define the (outer) turning point \(r_t>r_h\) by 
\begin{equation}
\omega^2=V_{\rm eff}(r_t)\,.
\label{eq:turning_points_WKB_rewrite}
\end{equation}
Using \(dr_*/dr=R^2/(\alpha\beta)\), therefore the Bohr--Sommerfeld quantization condition for the symmetric well becomes
\begin{equation}
2\int_{r_h}^{r_t} \frac{R^2\,dr}{\alpha(r)\beta(r)}
\sqrt{\omega_n^2-V_{\rm eff}(r)}
=
\pi\left(n+\frac12\right)\,.
\label{eq:BS_r_rewrite}
\end{equation}
For any given $n$, the classically allowed region is defined by \(\omega^2>V_{\rm eff}(r)\), with a turning point \(r_t>r_h\) determined by \eqref{eq:turning_points_WKB_rewrite}.  We can now use \eqref{eq:BS_r_rewrite} to generate a discrete WKB spectra \(\{\omega_{n,k}\}\) to be fed directly into spectral diagnostics. Given a discrete spectrum \(\{\omega_a\}\) (here \(a\) stands for the combined index \((n,k)\) over the selected modes), we will be able to define a Euclidean partition function $Z(\beta_E,t)$ and consequently the spectral form factor (SFF) $g(\beta_E,t)$:
\begin{equation}
Z(\beta_E,t)=\sum_a e^{-(\beta_E+it)\omega_a}\,,
\qquad
g_{\beta_E}(t)=g(\beta_E,t)=\frac{|Z(\beta_E,t)|^2}{|Z(\beta_E,0)|^2}\,.
\label{eq:SFF_defs_rewrite}
\end{equation}
Here the Euclidean inverse temperature \(\beta_E\) is analytically continued to complex time \(\beta_E+it\). These will be our quantities of interest in section \ref{sec:numerics_SFF_3D}. 

At this point, it is important to note that a WKB analysis also appeared for the probe dynamics in the DS$_3$  background in \cite{Solodukhin:2005qy}. There the author solved for the WKB spectrum with an approximated effective potential, and obtained a simple \(\lambda\) dependent mass gap along with a tower of frequencies separated trivially. However, here we'll numerically implement the Bohr-Sommerfeld condition for the exact potential, and obtain a non-trivial spectra.

\section{Numerical WKB spectrum and the spectral form factor}
\label{sec:numerics_SFF_3D}

The spectral form factor (SFF) provides a particularly sharp probe of long-time spectral correlations, and in gravitational settings it has become one of the standard diagnostics for distinguishing coarse discreteness from genuinely correlated spectra. In random-matrix ensembles, the canonical dip--ramp--plateau (DRP) structure reflects three regimes: an early-time decay (dip), an intermediate-time growth controlled by pair correlations (ramp), and a late-time saturation set by the finite effective dimension of the spectrum (plateau). In the modern gravity literature, this structure was emphasized in the black-hole/random-matrix comparison of \cite{Cotler:2016fpe} and then sharpened in near-AdS\(_2\)/JT gravity, where the ramp and plateau are tied to non-perturbative effects and ensemble interpretations of the gravitational path integral \cite{Cotler:2016fpe,Maldacena:2016upp,Saad:2019lba,Stanford:2019vob,Bhattacharyya:2023gvg}.\footnote{In the present DS$_3$ wormhole setup, the SFF should first be understood as a diagnostic of finite-throat discreteness and late-time phase correlations of the probe spectrum. A dip–ramp–plateau like morphology can arise already from a finite, structured normal-mode spectrum, and therefore should not by itself be identified with having a random matrix theory origin.}
In our current work, we would not have much to say about such path integrals, but rather our goal will be to investigate probe dynamics in our DS$_3$ wormhole setup. Once the throat regulator produces a discrete probe spectrum, the SFF becomes the natural quantity to test whether the late-time data exhibit the expected DRP morphology. Note that in our study the DRP is obtained in the probe spectrum. As mentioned earlier, the geometry here acts as a fixed Lorentzian background for the scalar probe. In particular, it is not a Euclidean replica wormhole, and the SFF studied below is not obtained from a gravitational path integral or a sum over bulk topologies. The key point is that the wormhole parameter \(\lambda\) plays a role analogous (at the level of diagnostics) to stretched-horizon/brick wall regulators \cite{tHooft:1985ys}, while retaining a geometric interpretation in the bulk. In doing so, we hope to emphasize the possible importance of such near-horizon modified geometries, as non-perturbative saddles. Some recent works in this latter direction (in the context of brick wall) appear in \cite{Das:2022wcj,Das:2023btz,Das:2023simp,Das:2023frm,Krishnan:2023nms,Terashima:2026fix}, and a central lesson from these works is that a near-linear ramp can emerge rather robustly in probe spectra, even before one has established full random-matrix universality in the strongest sense. We will compare our results to their analysis at the end of this section.

\subsection{Numerical level scans}
\label{subsec:Veff_analysis}

As we have seen in section \ref{subsec:WKB_quantization_rewrite}, the effect of \(k\) on $V_{\rm eff}$ comes from the angular-momentum term, which makes the potential steeper as \(|k|\) increases. This shifts the turning points and changes the WKB phase integral. As a result, the energy levels arrange themselves into \(k\)-dependent families \(\omega_{n,k}\). After coarse-graining, this structure can produce longer intermediate-time ramps in the spectral form factor, in line with the mode-organization picture found in BTZ brick wall studies \cite{Das:2022wcj,Das:2023simp,Krishnan:2023nms}.
Increasing \(n\) raises the WKB action by \(\pi\) per unit \(n\), which generically pushes \(\omega_{n,k}\) upward and increases the number of participating levels in a fixed frequency window. These dependencies are made explicit in the level scans below in figures \ref{fig:omega_vs_k.pdf} and \ref{fig:omega_vs_n}. Note that for larger \(n\), \(\omega\) values start from higher corresponding \(k\) quantum numbers. One can already see from these scans that the WKB levels are not random.  They organize into smooth $k$-dependent and $n$-dependent bands unlike what happens for RMT.\footnote{We thank Pratik Rath for a discussion on this point.} As a result, within a fixed band, this structured spectrum can produce coherent dephasing and a ramp-like recovery. However, different radial bands remain independent normal-mode families, and their combined contribution generally obscures rather than improving the approximate linearity of the lowest-band ramp.

\begin{figure}[h]
\centering
\begin{subfigure}[t]{0.49\textwidth}
  \centering
  \includegraphics[width=\textwidth]{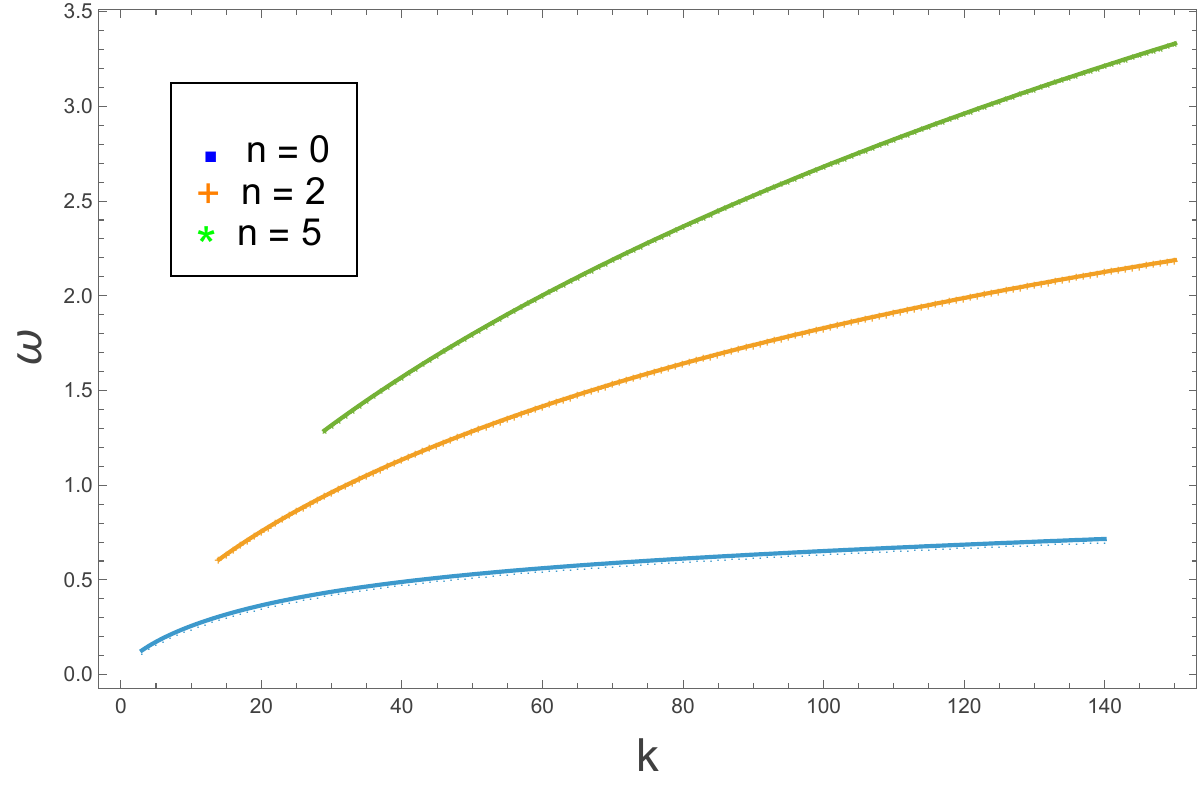}
  \caption{\(\omega_{n,k}\) vs.\ \(k\) for several fixed \(n\).}
  \label{fig:omega_vs_k.pdf}
\end{subfigure}\hfill
\begin{subfigure}[t]{0.49\textwidth}
  \centering
  \includegraphics[width=\textwidth]{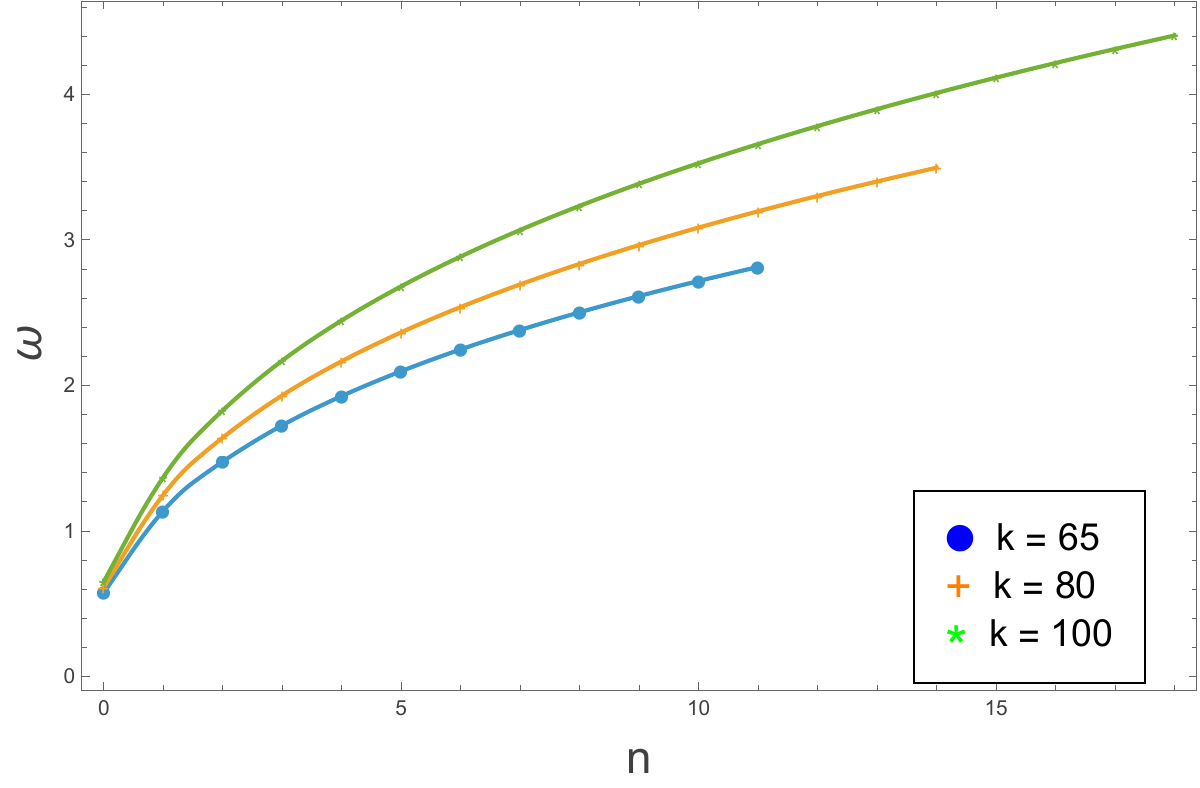}
  \caption{\(\omega_{n,k}\) vs.\ \(n\) for several fixed \(k\).}
  \label{fig:omega_vs_n}
\end{subfigure}
\caption{Dependence of WKB levels on \(k\) and \(n\), computed by solving the
Bohr--Sommerfeld equation \eqref{eq:BS_r_rewrite}.}
\label{fig:omega_nk_dependence}
\end{figure}

To obtain the above plots, we first solved for the turning points $r_t(\omega)$ given by $\omega^2-V_{\rm eff}(r_t)=0$. Upon inserting this turning point solution back into
\begin{equation}
2\int_{r_h}^{r_t}\frac{R^2\,dr}{\alpha(r)\beta(r)}
\sqrt{\omega_n^2-V_{\rm eff}(r)}
=\pi\left(n+\frac12\right)\,,
\label{eq:BS_repeat}
\end{equation}
we can now numerically solve for the discrete set of frequencies $\omega_{n,k}$ (for chosen sets of each angular momentum \(k\) and radial band \(n\). We also have \(r_t>r_h\)). This discrete set of $\omega_{n,k}$ is truncated in practice by finite windows \(|k|\le k_{\max}\) and \(0\le n\le n_{\max}\). Some discussions of these numerical techniques which lead to figure \ref{fig:omega_nk_dependence} onwards, are given in appendix \ref{subsec:WKB_data}. 

\subsection{WKB spectrum and SFF analysis}
\label{subsec:wkb_spectrum_discussion}

\begin{figure}
\centering
\includegraphics[width=0.7\textwidth]{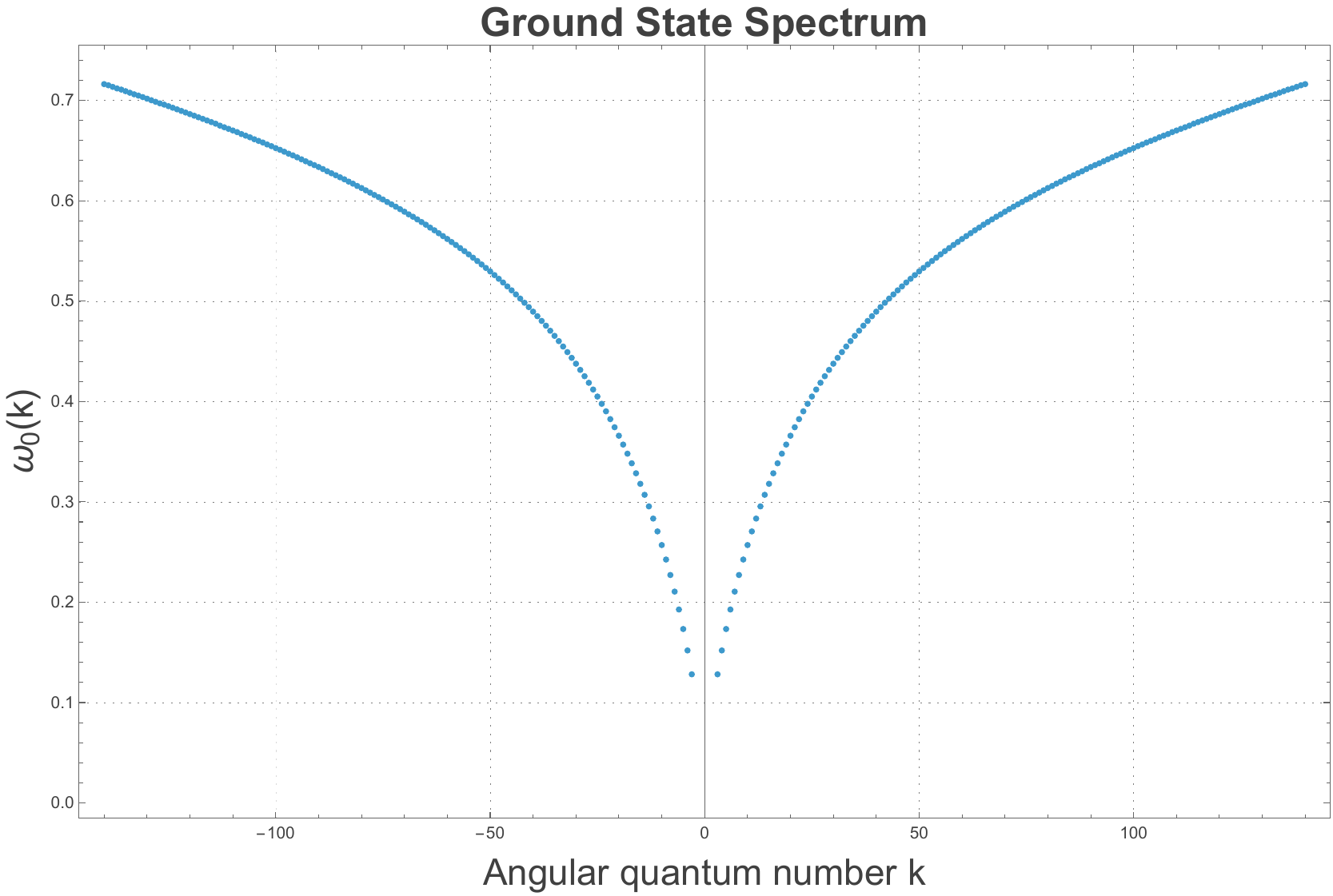}
\caption{WKB spectrum: lowest-band frequencies \(\omega_0(k)\) obtained from the Bohr--Sommerfeld condition
\eqref{eq:BS_r_rewrite} with \(n=0\), for \(k\in[-150,150]\) at \((R,r_h,\lambda)=(1,3,10^{-4})\).}
\label{fig:wkb_spectrum_omega0}
\end{figure}

Implementing the numerical technique discussed in the previous subsection, we obtain the   ground state (\(n=0\)) WKB spectrum \(\omega_0(k)\) as given by figure \ref{fig:wkb_spectrum_omega0}. The key structural point is discreteness: for \(\lambda\neq 0\), the DS wormhole throat smoothens the would-be horizon at \(r=r_h\) and yields a discrete spectrum without an imposed stretched-horizon boundary condition. For an equivalent finding in the presence of the stretched horizons, see \cite{Das:2023btz}. 

\begin{figure}
\centering
\begin{subfigure}[t]{0.49\textwidth}
  \centering
  \includegraphics[width=\textwidth]{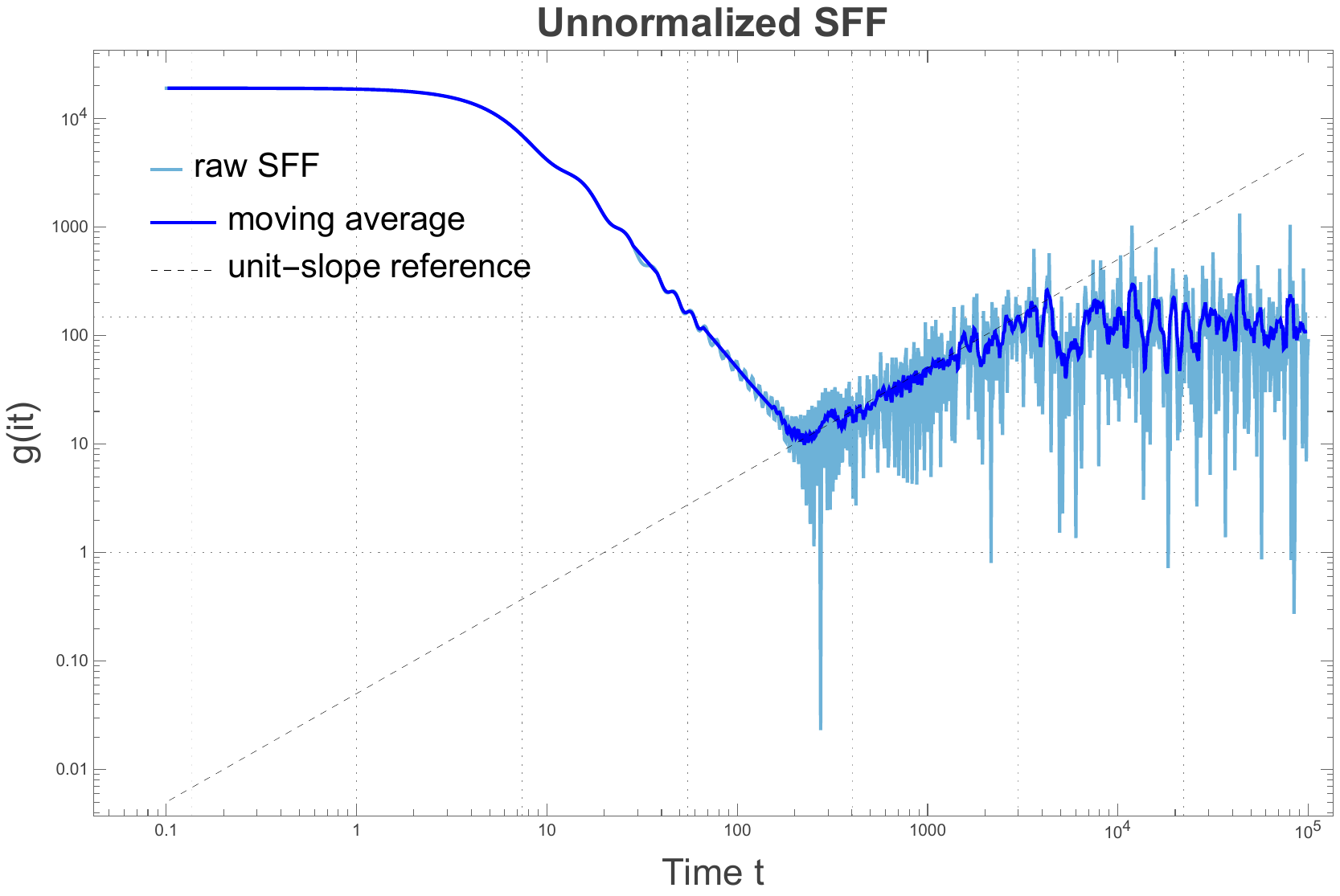}
  \caption{Unnormalized.}
  \label{fig:sff_unnorm}
\end{subfigure}\hfill
\begin{subfigure}[t]{0.49\textwidth}
  \centering
  \includegraphics[width=\textwidth]{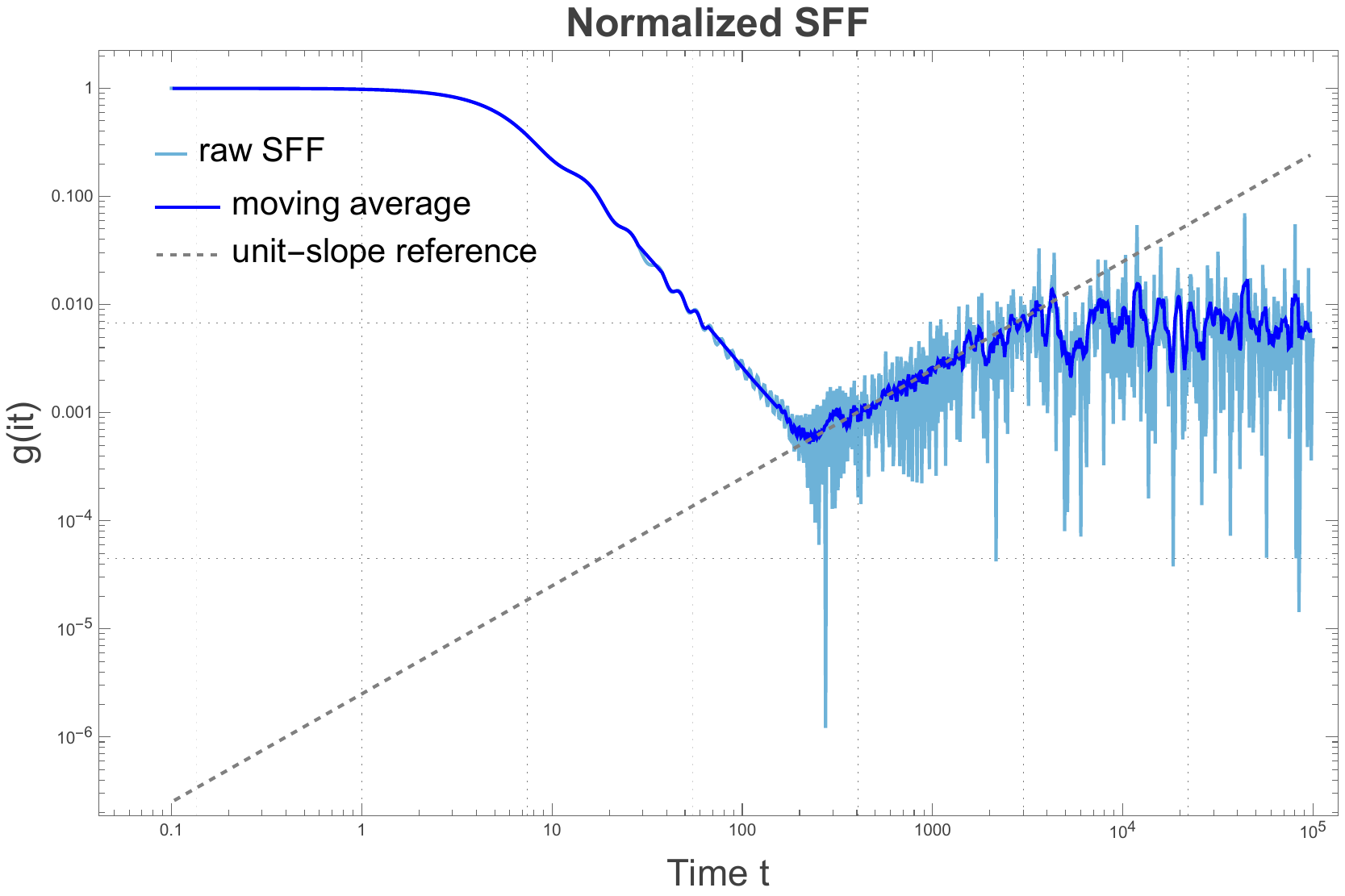}
  \caption{Normalized.}
  \label{fig:sff_norm}
\end{subfigure}
\caption{\(\beta_E=0\) SFF computed for \((n,R,r_h,\lambda)=(0,1,3,10^{-4})\) and \(k\in[0,150]\). The dashed reference line has slope \(1\) and is anchored through the ramp to facilitate visual comparison to linear-ramp expectations \cite{Cotler:2016fpe,Das:2022wcj,Das:2023btz}. The value on the vertical axis distinguishes between unnormalized and normalized cases.}
\label{fig:sff_loglog_ramp}
\end{figure}

\subsubsection{{$\beta_E=0$} SFF: dip--ramp--plateau}
\label{subsec:beta0_sff_discussion}

Given the discrete spectrum \(\{\omega_j\}\) (with \(j\) a collective index, e.g.\ \(j=k\) when restricting to \(n=0\)), we compute the SFF for $\beta_E=0$. In that case, we have
\begin{equation}
Z(it)=\sum_j e^{-it\omega_j},
\qquad
g_0(t)=g(it)=\frac{|Z(it)|^2}{|Z(0)|^2}.
\label{eq:SFF_beta0_repeat}
\end{equation}
Figure~\ref{fig:sff_loglog_ramp} displays the resulting \(\beta_E=0\) SFF. The normalized plot emphasizes the DRP profile: early-time decay (dip), an intermediate-time ramp, and a late-time plateau. A unit-slope reference line is included as a guide, since slope \(\sim 1\) ramps are the canonical expectation in random matrix theory (RMT) style diagnostics and are observed in BTZ brick wall settings after appropriate coarse-graining/averaging at \(\beta_E=0\) \cite{Cotler:2016fpe,Das:2022wcj,Das:2023btz}. We have also placed a moving average for the plotted SFF where the raw SFF has been averaged over a small time-window, so that the coarse DRP profile becomes cleaner. 

\begin{figure}
\centering
\begin{subfigure}[t]{0.49\textwidth}
  \centering
  \includegraphics[width=\textwidth]{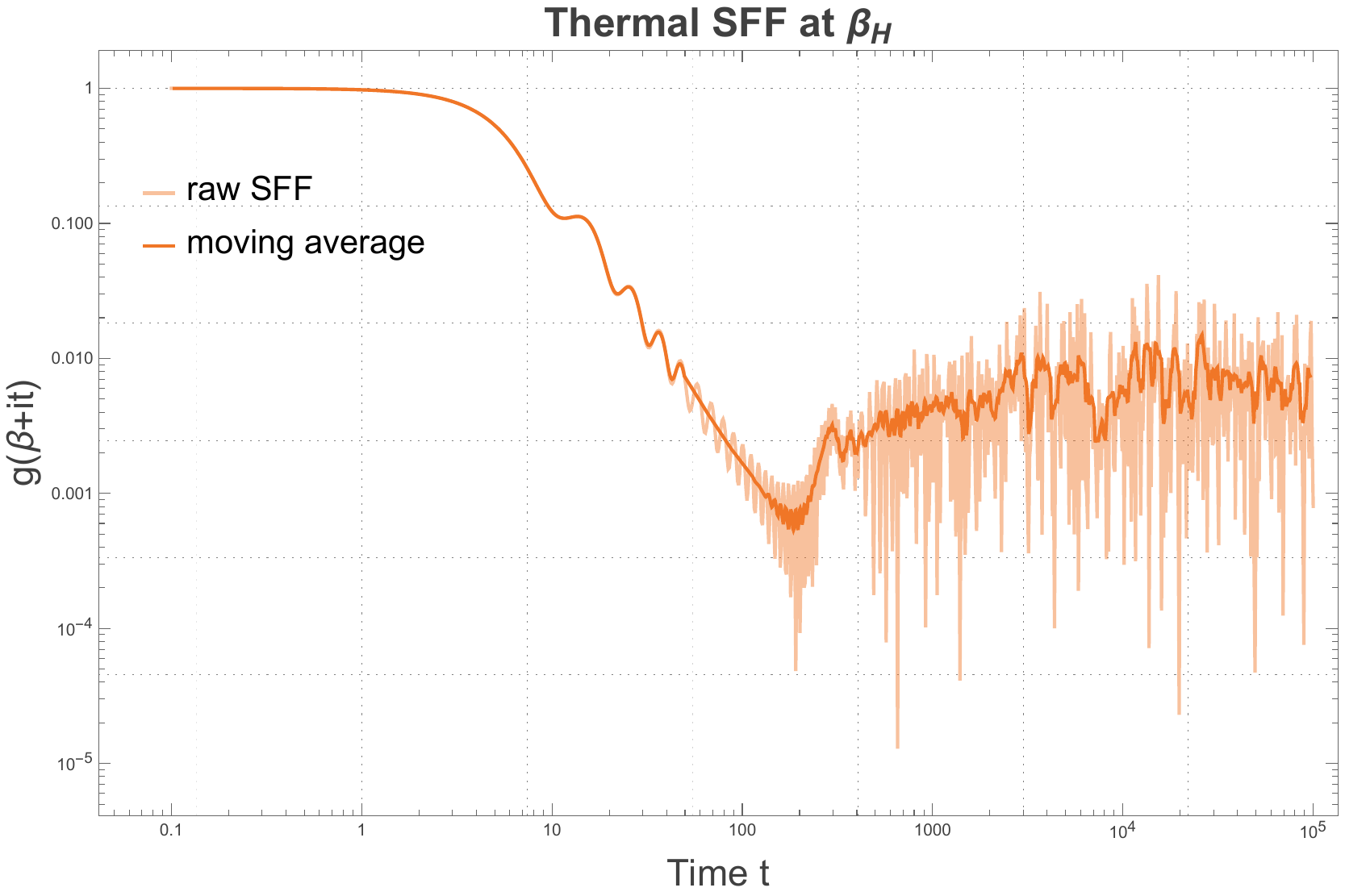}
  \caption{Normalized.}
  \label{fig:sff_norm_thermal_ground}
\end{subfigure}\hfill
\begin{subfigure}[t]{0.49\textwidth}
  \centering
  \includegraphics[width=\textwidth]{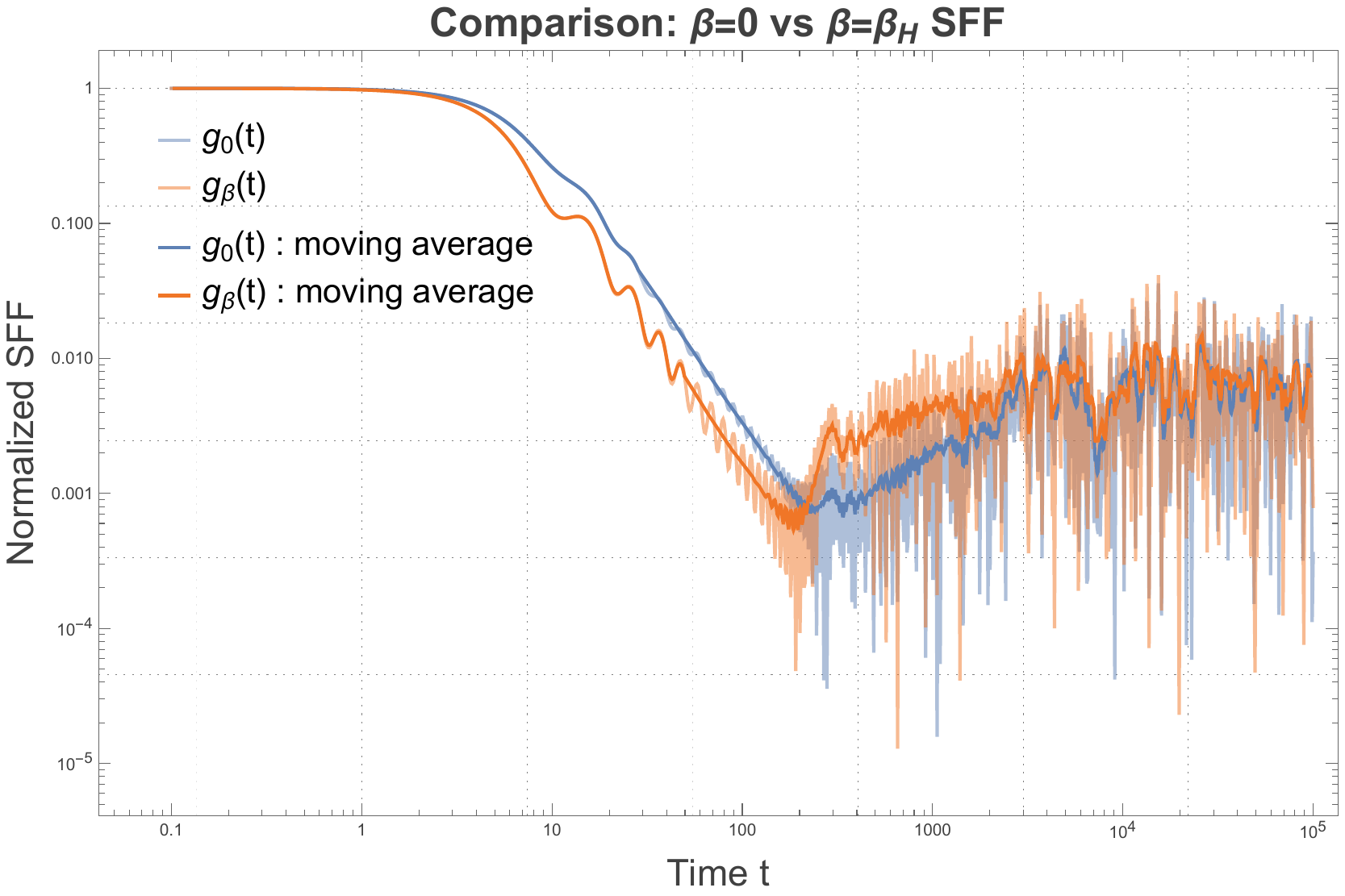}
  \caption{Comparison.}
  \label{fig:sff_thermal_vs_beta0}
\end{subfigure}
\caption{Thermal SFF for the ground band (\(n=0\)) with \(k\in[0,350]\).}
\label{fig:sff_thermal}
\end{figure}

At finite \(\beta_E\), Boltzmann weighting suppresses higher-frequency modes and reduces sensitivity to the UV end of the truncated spectrum. One expects the same qualitative DRP sequence but with shifted time scales. Figure \ref{fig:sff_thermal} shows the normalized thermal SFF for the ground band and compares it against the \(\beta_E=0\) case. From the figure, we notice that \(\beta_E=\beta_H\) spoils the linearity of the ramp. In particular, the orange curve shows fluctuations at the beginning of the ramp, and deviates from the unit slope of the blue curve.

\begin{figure}
\centering
\begin{subfigure}[t]{0.49\textwidth}
  \centering
  \includegraphics[width=\textwidth]{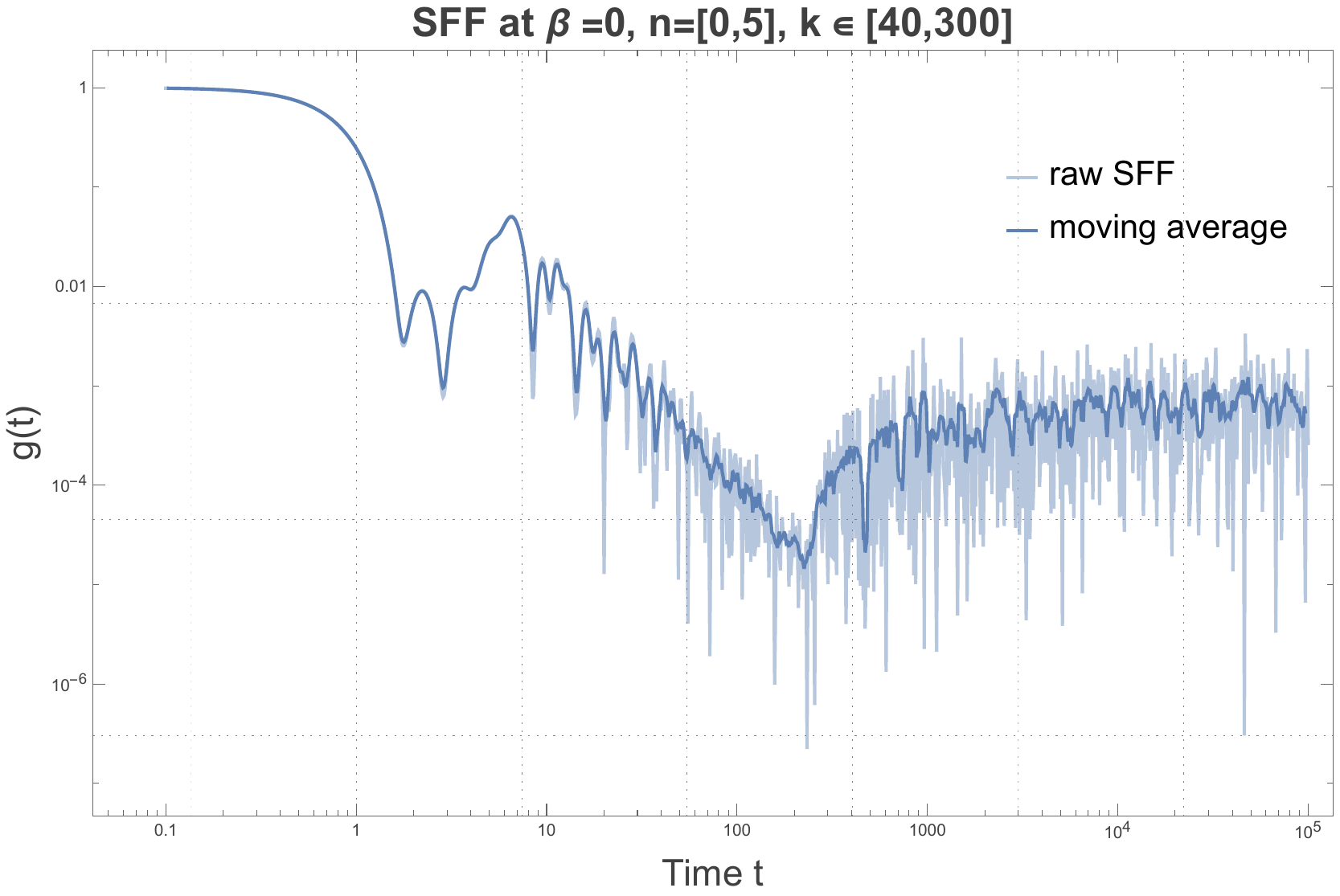}
  \caption{SFF at \(\beta_E=0\).}
  \label{fig:sff_nsum_beta0}
\end{subfigure}\hfill
\begin{subfigure}[t]{0.49\textwidth}
  \centering
  \includegraphics[width=\textwidth]{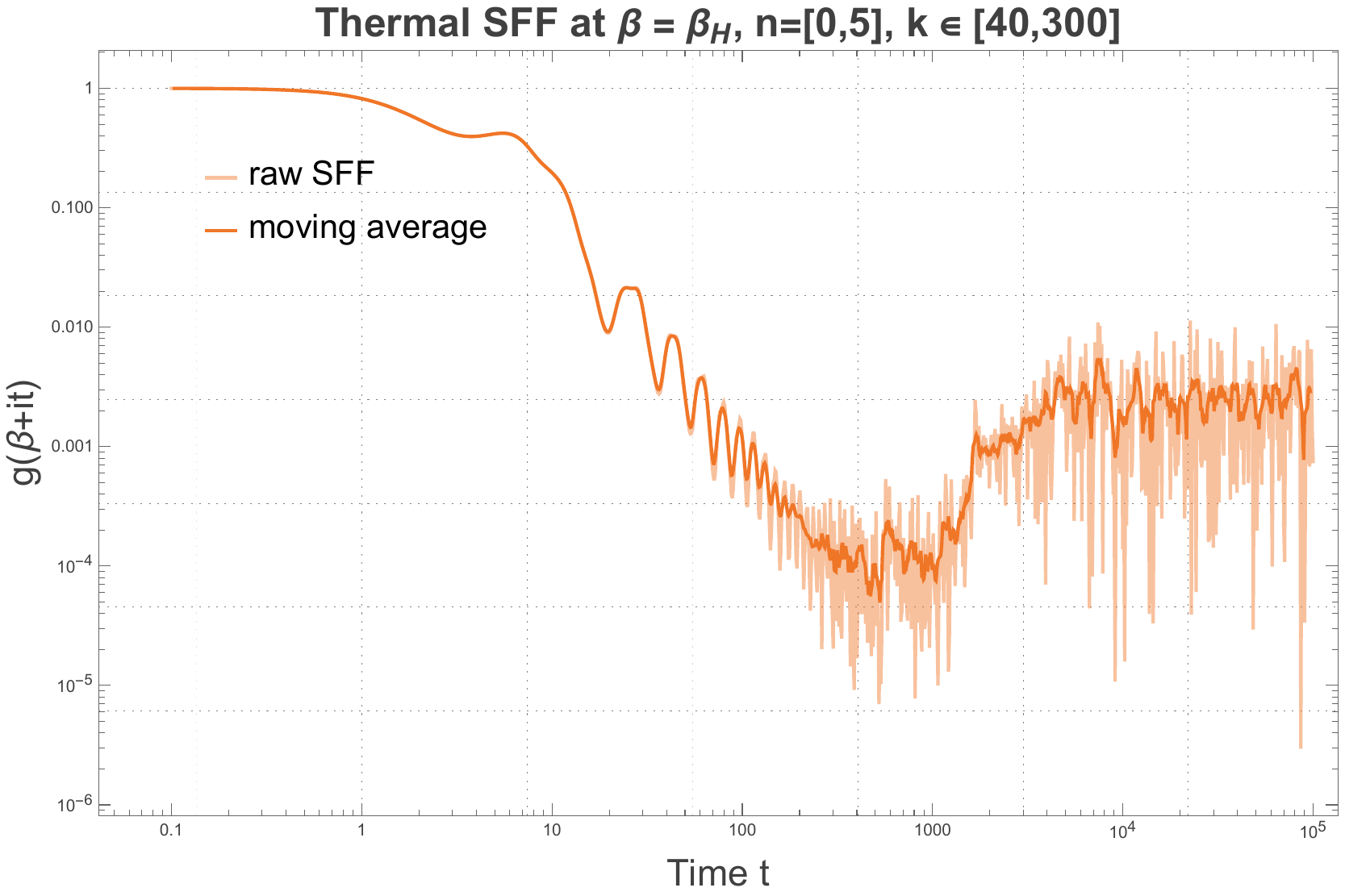}
  \caption{SFF at \(\beta_E=\beta_H\).}
  \label{fig:sff_nsum_betaH}
\end{subfigure}
\caption{SFF summed over \(n\in[0,8]\) with \(k\in[40,300]\).}
\label{fig:sff_thermal_n_m_summed}
\end{figure}

\subsubsection{Including multiple radial bands}
\label{subsec:multi_band_sff_discussion}

One may also include multiple radial bands to investigate the resulting SFF structure. Figure \ref{fig:sff_thermal_n_m_summed} shows SFFs obtained by summing over \(n\in[0,8]\) with a larger angular window \(k\in[40,300]\). From this figure, we see that the SFF retains a DRP profile, but it does not exhibit a clean linear ramp as found in the single band case. Curiously in our case, upon averaging over $(n,k)$, the ramp region deviates from a clean, linear slope. It could be because there are missing \(\omega_{n,k}\) for lower values of \(k\) and corresponding to higher \(n\) (as is clear from figure \ref{fig:omega_nk_dependence}). It may also be the result of the fact that with averaging over $(n,k)$, the previously present level pair correlations are modified.

\subsubsection{\texorpdfstring{$\lambda$}{lambda}-ensemble averaging}
\label{subsec:lambda_ensemble_discussion}

Fixed-\(\lambda\) SFFs can exhibit non-universal oscillations. A standard way to isolate the coarse DRP envelope is to average over a narrow distribution of a control parameter, mirroring the stretched-horizon averaging used in brick wall analyses \cite{Das:2022wcj,Das:2023btz}. In the DS-wormhole, \(\lambda\) is an intrinsic geometric parameter controlling the throat microstructure, so \(\lambda\)-averaging is a natural geometric analog of parameter averaging. We choose a Gaussian ensemble for $\lambda$ in the same spirit of the above mentioned stretched horizon program as a micro-canonical ensemble of non-perturbatively deformed states around BTZ blackhole.

\begin{figure}[h]
\centering
\includegraphics[width=0.7\textwidth]{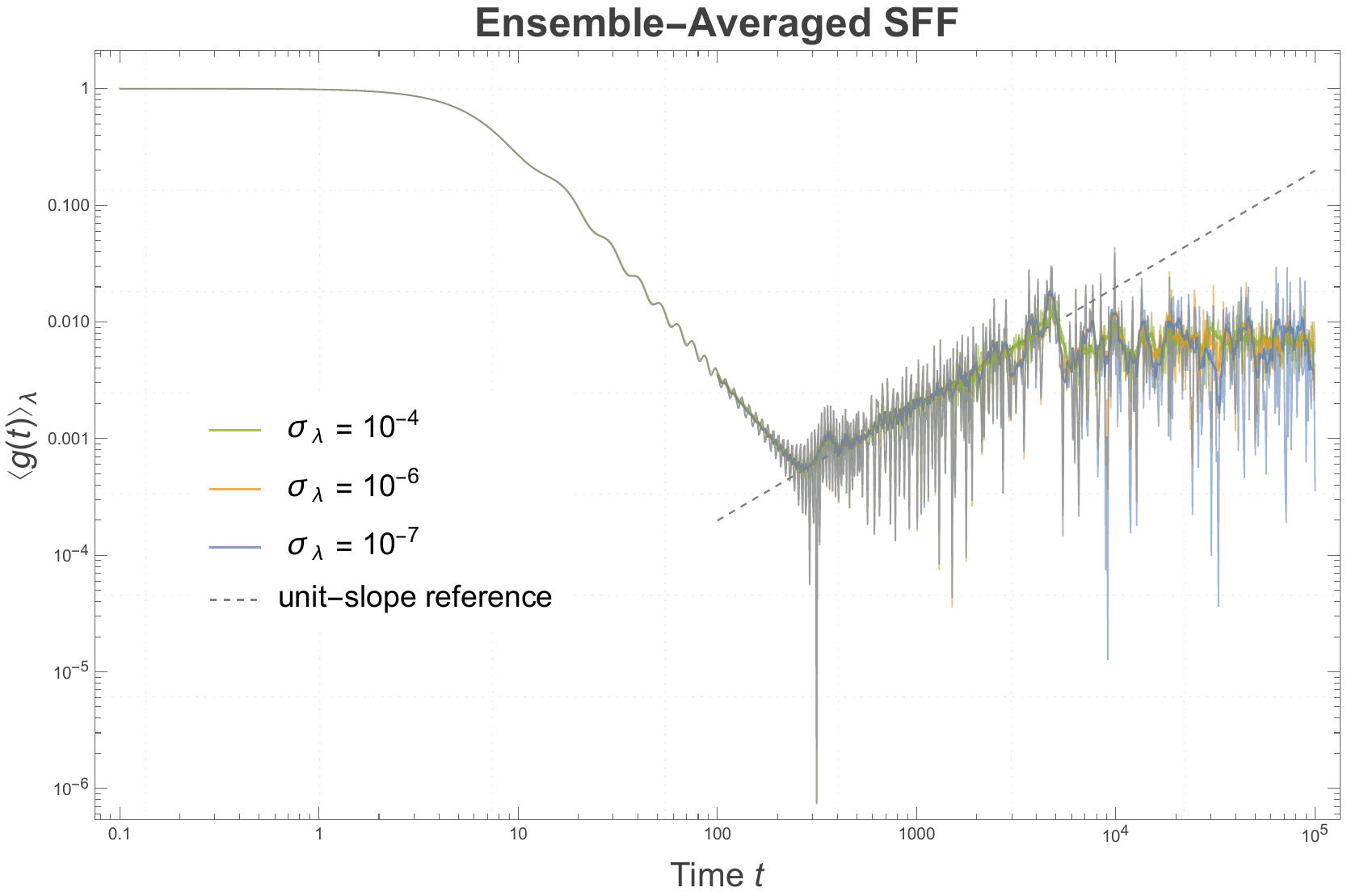}
\caption{Ensemble-averaged diagnostics at \(n=0\), for \(k\in[0,350]\) with
\((R,r_h,\lambda_0)=(1,3,5\times10^{-4})\). A unit-slope reference line is placed for visualizing ramp linearity. We have also added three different choices of $\lambda$-variance, in each case with 10 samples. $\sigma_\lambda\sim\mathcal{O}(\lambda)$ (green) suppresses the raw fluctuation the most, although the suppression depends both on the sample size and the $\sigma_\lambda$.}
\label{fig:ensemble_avg}
\end{figure}

Figure~\ref{fig:ensemble_avg} shows the ensemble-averaged SFF at \(n=0\) for a Gaussian distribution of \(\lambda\) centered at some \(\lambda_0\). 
Averaging suppresses spiky fluctuations and makes the ramp segment substantially cleaner, consistent with the linear-ramp phenomenology extracted from regulated BTZ normal modes \cite{Das:2022wcj,Das:2023simp,Krishnan:2023nms}. After this coarse graining the ramp becomes strikingly linear for the DS wormholes.

In summary: (i) \(\lambda\neq 0\) produces a confining effective potential and a discrete WKB spectrum without an imposed stretched-horizon boundary condition; (ii) the resulting SFF exhibits a clear DRP profile with an extended near-linear ramp; (iii) thermal weighting modifies the time scales and spoils the linearity of the ramp, but preserves the qualitative DRP structure; and (iv) \(\lambda\)-ensemble averaging suppresses non-universal oscillations and exposes the ramp more cleanly, paralleling parameter-averaging practices in BTZ brick wall analyses \cite{Das:2022wcj,Das:2023btz,Krishnan:2023nms}. 

\subsection{The linear ramp and BTZ/brick wall results} \label{subsec:brickwall_comparison}

A central lesson from recent BTZ/brick wall and stretched-horizon studies is that a near-linear ramp can emerge rather robustly in probe spectra, even before one has established full random-matrix universality in the strongest sense. In particular, recent works have clarified that a linear ramp can arise in deterministic spectra (e.g.\ logarithmic spectra), and that the ramp and level repulsion diagnose different aspects of the spectrum---long-range versus short-range correlations \cite{Das:2022wcj,Das:2023simp,Das:2023btz}. This is directly relevant for our analysis: the approximately unit-slope ramp seen in the normalized SFF is a meaningful indicator of nontrivial late-time spectral organization, but by itself should not be interpreted as a proof of full RMT behavior. Rather, in line with the above literature, our fixed-\(\lambda\) and \(\lambda\)-averaged diagnostics should be viewed as complementary probes: fixed-\(\lambda\) data retain model-specific oscillatory structure, while \(\lambda\)-averaging suppresses non-universal fluctuations and makes the coarse DRP envelope (especially the ramp-to-plateau crossover) more transparent. In this sense, our DS-wormhole construction further expands the BTZ/brick wall story by realizing the regulator geometrically and reproducing the same qualitative SFF phenomenology expected for finite-\(N\) black-hole microphysics.

In the present DS-wormhole, discreteness arises without imposing an auxiliary boundary condition (as already emphasized): \(\lambda\neq 0\) replaces the would-be horizon by a smooth throat at \(r=r_h\), yielding a confining effective potential in the Schr\"odinger problem. Nevertheless, at the level of diagnostics the two setups are directly comparable: both admit (i) a discrete spectrum, (ii) a tunable geometric control parameter (stretched-horizon location vs.~\(\lambda\)), and (iii) DRP structure in an appropriately coarse-grained/averaged SFF. This makes the DS throat a purely geometric analog of the brick wall regulator, and motivates comparing ramp linearity and plateau stabilization to the BTZ results \cite{Das:2022wcj,Das:2023btz,Krishnan:2023nms}.

\section{Conclusions}
\label{sec:discussion_outlook}

This work was motivated by a tension between two standard expectations. On one hand, the perturbative HKLL reconstruction gives bulk fields as boundary-smeared operators with support that can extend to arbitrarily late boundary times as the bulk point approaches a horizon. On the other hand, at finite $N$ one expects a finite number of independent commuting observables in any finite bulk region, and hence a breakdown of the naive semiclassical operator map at sufficiently large time separations. 
Truncating the HKLL integral is equivalent to subtracting an ``excised'' contribution that localizes near the boundary at late times. 
We have demonstrated that the DS wormhole geometries furnish an equivalent non-perturbative modifications of the bulk geometry, where the non-perturbative corrections enter via the regulated wormhole throat parameter.  From the bulk perspective, the `regulator' is simply the absence of an infinite optical depth: the wave equation never allows propagation into an infinitely deep near-horizon region. In this sense, a throat can be viewed as a bulk completion of the cut-off/excision picture. We have made the equivalence between these two approaches completely explicit for throat-corrected AdS$_2$ black holes. In the context of AdS$_3$ black holes, we have provided a similar order of estimate for the deviation of bulk semiclassicality.

In three dimensions however, the separated radial equation is structurally richer and reduces to a general Heun equation. We emphasized two complementary representations of the same problem: (i) the exact Heun reduction, which cleanly characterizes the deformation compared to semiclassical bulk reconstruction (with respect to BTZ), and (ii) the Schr\"odinger reformulation with an explicit effective potential, which is better adapted to semiclassical quantization and spectral statistics. In order to study the non-perturbative aspects, the principal diagnostic used here is the spectral form factor (SFF), which we have studied both at fixed throat parameter $\lambda$, and by considering an average over $\lambda$. Fixed-$\lambda$ data retain model-specific oscillatory features, whereas $\lambda$-averaging suppresses these oscillations and more directly reveals the dip--ramp--plateau envelope associated with level correlations. This parallels the logic of stretched-horizon/brick wall analyses, while keeping the regulator geometric rather than an imposed boundary condition at an auxiliary surface. It is interesting to see that this sort of wormhole geometries can reproduce the expectations from brick wall/stretched horizon like effective theories. However, in order to probe into the true chaotic structure of a theory of quantum gravity one needs to further deform these geometries via adding further deformations to the effective potential and study the unfolded spectral properties. This could be an interesting future direction to explore.

\bigskip
\goodbreak
\centerline{\bf Acknowledgements}
\noindent
We would like to thank Dan Kabat and Sergey Solodukhin for discussions, and their valuable feedback on the manuscript. We also would like to thank Abhay Kumar Singh for an initial collaboration on this project. 
The work of MN is supported by DST Inspire grant and the work of DS is partly supported by DST-FIST grant SR/FST/PSI-225/2016 and CRG grant CRG/2023/000904. MN has taken assistance from chatGPT in writing Mathematica codes for some of the plots appearing in this paper.

\appendix

\section{Bulk dynamics in DS$_2$ wormholes}
\label{subsec:rstar_AdS2}

The 2D wormhole model is analytically tractable and serves to make the geometric cut-off mechanism explicit. We introduce the tortoise coordinate
\begin{equation}
r_*=\int \frac{R^2}{\sqrt{(r^2-r_h^2)(r^2-r_h^2+\lambda^2)}}\,dr\,,
\label{eq:rstar_def}
\end{equation}
and fix the additive constant by requiring $r_*(r\to\infty)=0$.
Then $r_*(r)<0$ for finite $r$ and it decreases monotonically as $r\to r_h^+$. 
The minimum (or the maximum, if working with absolute values) value of the tortoise coordinate is
\begin{equation}
r_*^{\max}:=\lim_{r\to r_h^+}r_*(r)
=-\int_{r_h}^{\infty}
\frac{R^2}{\sqrt{(r^2-r_h^2)(r^2-r_h^2+\lambda^2)}}\,dr.
\label{eq:rstarmax_def}
\end{equation}
Therefore a key geometric difference of DS$_2$ wormhole from a black hole is that $r_*^{\max}$ is finite for $\lambda\neq 0$.
The above equation \eqref{eq:rstarmax_def} can be solved exactly to obtain 
\begin{equation} 
r_*^{\max}=-\frac{R^2}{r_h}\,K\!\left(\sqrt{1-\frac{\lambda^2}{r_h^2}}\right)\,,
\label{eq:rstarmax_exact}
\end{equation}
which yields 
\begin{equation}
r_*^{\max}\approx -\frac{R^2}{r_h}\log\left(\frac{4r_h}{\lambda}\right)
\qquad \text{for}\qquad \lambda\ll r_h
\label{eq:rstarmax_log}
\end{equation}
to the leading order in $\frac{\lambda}{r_h}$ expansion.
In the case for a black hole, $r_*^{\max}\to-\infty$ with $\lambda\to 0$.
Using these coordinates one can show that the massless Klein-Gordon equation is given by 
\begin{equation}
\Box\phi=\frac{1}{\Omega^2}\left(-\partial_t^2+\partial_{r_*}^2\right)\phi=0,
\qquad
\text{with}
\qquad
\Omega^2(r)=\frac{r^2-r_h^2+\lambda^2}{R^2}\,.
\label{eq:KG_conformal}
\end{equation}

We have already discussed bulk reconstruction for the massless case in section \ref{sec:AdS2_DSW_cutoff}, and for reasons mentioned in footnote \ref{fn:massless}, it is also where the non-locality can be estimated precisely in terms of another boundary CFT operator. However for completeness, let us now briefly present the corresponding analysis for a massive scalar. Starting from the Klein--Gordon equation
\begin{equation}
(\Box-m^2)\Phi=0
\label{eq:KG_massive}
\end{equation}
with the mode ansatz 
\begin{equation}
\Phi_{\omega}(t,r)=e^{-i\omega t}\,\mathcal R_\omega(r)\,,
\label{eq:mode_ansatz_massive}
\end{equation}
one obtains
\begin{equation}
(r^2-r_h^2)\,\mathcal R_\omega''(r)
+\frac{r(2r^2-2r_h^2+\lambda^2)}{r^2-r_h^2+\lambda^2}\,\mathcal R_\omega'(r)
+\left(
\frac{R^4\omega^2}{r^2-r_h^2+\lambda^2}-m^2R^2
\right)\mathcal R_\omega(r)=0\,.
\label{eq:radial_eq_r_massive}
\end{equation}\\
It is convenient to introduce the coordinates
\begin{equation}
z=\frac{r_h^2}{r^2}
\qquad \text{and} \qquad
z_s=\frac{r_h^2}{r_h^2-\lambda^2}\,,
\label{eq:z_zs_def_massive}
\end{equation}
so that the right exterior patch is given by $0<z<1$ (with $z\to 0$ being the asymptotic AdS boundary, and $z\to 1$ is the classical horizon). 
The resulting equation is 
\begin{equation}
\mathcal R_\omega''(z)
+
\left(
\frac{1}{2z}
+\frac{1}{2(z-1)}
+\frac{1}{2(z-z_s)}
\right)\mathcal R_\omega'(z)
+
\frac{\frac{z_sR^4\omega^2}{r_h^2}z+R^2m^2(z-z_s)}
{4z^2(z-1)(z-z_s)}
\,\mathcal R_\omega(z)=0\,.
\label{eq:fuchsian_massive}
\end{equation}\\
Near the AdS boundary $z=0$, taking $\mathcal R_\omega(z)\sim z^{\alpha_0}$ gives the indicial
equation
\begin{equation}
4\alpha_0(\alpha_0-1)+2\alpha_0-R^2m^2=0\,,
\label{eq:indicial_massive}
\end{equation}
with roots
\begin{equation}
\alpha_\pm=\frac{1\pm\sqrt{1+4R^2m^2}}{4}\,.
\label{eq:alpha_pm_massive}
\end{equation}
These are precisely the (non-)normalizable modes associated to the extrapolate dictionary
\begin{equation}
\alpha_+=\frac{\Delta}{2}\,,
\qquad
\alpha_-=\frac{1-\Delta}{2}\,.
\label{eq:alpha_pm_Delta_massive}
\end{equation}
Focusing on the normalizable branch 
\begin{equation}
\mathcal R_\omega(z)=z^{\Delta/2}y_\omega(z)\,,
\label{eq:R_equals_zD_y}
\end{equation}
and substituting \eqref{eq:R_equals_zD_y} into \eqref{eq:fuchsian_massive}, one obtains
\begin{equation}
y_\omega''(z)
+
\left(
\frac{\Delta+\frac12}{z}
+\frac{1}{2(z-1)}
+\frac{1}{2(z-z_s)}
\right)y_\omega'(z)
+
\frac{\alpha\beta\,z-q_+}{z(z-1)(z-z_s)}\,y_\omega(z)=0\,,
\label{eq:heun_form_massive}
\end{equation}
where
\begin{equation}
\alpha=\alpha_+=\frac{\Delta}{2}\,,
\qquad
\beta=\frac{\Delta+1}{2}\,,
\qquad\text{and}\qquad
q_+=\frac{z_s\Delta^2+\Delta}{4}-\frac{R^4\omega^2}{4(r_h^2-\lambda^2)}\,.
\label{eq:heun_params_massive}
\end{equation}
Comparing \eqref{eq:heun_form_massive} with the standard general Heun equation, we identify the following behavior of the normalizable mode near the boundary
\begin{equation}
\mathcal R_\omega^{\rm norm}(z)
=
z^{\Delta/2}\,
\text{HeunG}\!\left(
z_s,\,
q_+,\,
\frac{\Delta}{2},\,
\frac{\Delta+1}{2},\,
\Delta+\frac12,\,
\frac12;\,
z
\right)\,.
\label{eq:Heun_normalizable_massive}
\end{equation}
Since, $\text{HeunG}(a,q,\alpha,\beta,\gamma,\delta;0)=1$, one has
\begin{equation}
\mathcal R_\omega^{\rm norm}(z)\sim  z^{\Delta/2}\qquad \text{as}
\qquad z\to 0\,.
\label{eq:boundary_limit_Heun_massive}
\end{equation}
Thus we can formally write down the right-patch massive scalar as
\begin{equation}
\Phi_R(t,z)
=
\int_0^\infty \frac{d\omega}{2\pi}
\left[
e^{-i\omega t}\,
z^{\Delta/2}\,
\text{HeunG}\!\left(
z_s,\,
q_+,\,
\frac{\Delta}{2},\,
\frac{\Delta+1}{2},\,
\Delta+\frac12,\,
\frac12;\,
z
\right)\mathcal O_\omega
+\text{h.c.}
\right]\,.
\label{eq:HKLL_massive_formal}
\end{equation}
Using the boundary Fourier transform of the mode \(\mathcal{O}_\omega\) we can write
\begin{equation}\label{eq:HKLL_massive_formal2}
\Phi_R(t,z)=
	z^{\Delta/2}
    \int_{-\infty}^{\infty} dt'
    \int_{0}^{\infty}
    \frac{d\omega}{2\pi}\,
    e^{-i\omega(t-t')}\,\text{HeunG}\!\left(
z_s,\,
q_+,\,
\frac{\Delta}{2},\,
\frac{\Delta+1}{2},\,
\Delta+\frac12,\,
\frac12;\,
z
\right)\mathcal O(t')+\text{h.c}\,.
\end{equation}

\section{Scalar probe in DS$_3$ and deviation from BTZ-locality}
\label{subsec:non_locality_est}

We now derive the small-\(\lambda\) expansion of the bulk reconstruction kernel and show explicitly how the full field decomposes into a BTZ contribution plus a correction term due to the finite size of the throat. The HKLL kernel for the DS$_3$ wormhole is given by (see \eqref{eq:DS_smearing_kernel})
\begin{align}
    K_{\mathrm{DS}}
    =
    \frac{z^{\Delta/2}}{2\pi}
    \sum_{k\in\mathbb{Z}}
    \int_{-\infty}^{\infty}
    \frac{d\omega}{2\pi}\,
    e^{-i\omega(t-t')+ik(\varphi-\varphi')}
    \,\text{HeunG}\left(
        z_s,
        q_{+},
        \frac{\Delta}{2}+\frac{iRk}{2r_h},
        \frac{\Delta}{2}-\frac{iRk}{2r_h},
        \Delta,
        \frac{1}{2};
        z
    \right)\,.
    \label{eq:DS_smearing_kernel_start}
\end{align}
where
\begin{equation}
z_s
=
\frac{r_h^2}{r_h^2-\lambda^2}\,,
\quad
\alpha
=
\frac{\Delta}{2}
+
\frac{iRk}{2r_h}\,,
\quad \text{and}\quad
\beta
=
\frac{\Delta}{2}
-
\frac{iRk}{2r_h}\,.
\label{eq:alpha_beta_def}
\end{equation}
The corrected accessory parameter is
\begin{equation}
q_+
=
z_s
\left[
\frac{\Delta(\Delta-1)}{4}
+
\frac{R^2k^2}{4r_h^2}
\right]
+
\frac{\Delta}{4}
-
\frac{R^4\omega^2}{4(r_h^2-\lambda^2)}\,,
\label{eq:qplus_corrected}
\end{equation}
where we used \(R^2m^2=\Delta(\Delta-2) \). Introducing the dimensionless parameter $\varepsilon=\frac{\lambda^2}{r_h^2}$,
and expanding our result in orders of $\varepsilon$, we obtain 
\begin{align}
q_+&=\left[1+\varepsilon+O(\varepsilon^2)\right]A_{\Delta k}+\frac{\Delta}{4}-\frac{R^4\omega^2}{4r_h^2}-\varepsilon\frac{R^4\omega^2}{4r_h^2}+O(\varepsilon^2)\nonumber\\
&=A_{\Delta k}+\varepsilon A_{\Delta k}+\frac{\Delta}{4}-\frac{R^4\omega^2}{4r_h^2}-\varepsilon\frac{R^4\omega^2}{4r_h^2}+O(\varepsilon^2)\nonumber\\
&=q_+^{(0)}+\varepsilon q_+^{(1)}+O(\varepsilon^2)\,.
\label{eq:qplus_expansion_before_collecting}
\end{align}
Here we have used the shorthand
\begin{equation}
A_{\Delta k}
\equiv
\frac{\Delta(\Delta-1)}{4}
+
\frac{R^2k^2}{4r_h^2}\,.
\label{eq:ADeltak_def}
\end{equation}
The leading order piece is therefore
\begin{equation}
q_+^{(0)}
=
A_{\Delta k}
+
\frac{\Delta}{4}
-
\frac{R^4\omega^2}{4r_h^2}\,,
\label{eq:qplus0_def}
\end{equation}
and it can be shown that $q_+^{(0)}=q_{\rm BTZ}$. 
On the other hand, at first order in \(\varepsilon\), we have
\begin{equation}
q_+^{(1)}
=
A_{\Delta k}
-
\frac{R^4\omega^2}{4r_h^2}=\frac{\Delta(\Delta-1)}{4}
+
\frac{R^2k^2}{4r_h^2}
-
\frac{R^4\omega^2}{4r_h^2}\,.
\label{eq:qplus1_def_first}
\end{equation}
Using the smearing Kernel for the present case (as mentioned above) and defining 
\begin{equation}
\mathcal H(a,q;z)
\equiv
\operatorname{HeunG}
\left(
a,q,
\alpha,\beta,\Delta,\frac12;z
\right)\,,
\label{eq:Hcal_def}
\end{equation}
we can write 
\begin{equation}
K_{\rm DS}(z;\lambda)
=
\frac{z^{\Delta/2}}{2\pi}
    \sum_{k\in\mathbb{Z}}
    \int_{-\infty}^{\infty}
    \frac{d\omega}{2\pi}\,
    e^{-i\omega(t-t')+ik(\varphi-\varphi')}
\,\mathcal H
\left(
1+\varepsilon+O(\varepsilon^2),
q_{\rm BTZ}+\varepsilon q_1+O(\varepsilon^2);
z
\right)\,.
\label{eq:kernel_Hcal_eps}
\end{equation}
The final answer takes the form 
\begin{equation}
K_{\rm DS}(z;\lambda)
=
K_{\rm BTZ}(z)
+
\varepsilon K_{DS}^{(1)}(z)
+
O(\varepsilon^2)\,,
\label{eq:kernel_split_eps}
\end{equation}
with 
\begin{equation}
K_{\rm BTZ}(z)
=\frac{1}{2\pi}
    \sum_{k\in\mathbb{Z}}
    \int_{-\infty}^{\infty}
    \frac{d\omega}{2\pi}\,
    e^{-i\omega(t-t')+ik(\varphi-\varphi')}
z^{\Delta/2}
(1-z)^s
\,{}_2F_1
\left(
\frac{\Delta}{2}
+
s
+
\frac{iRk}{2r_h},
\frac{\Delta}{2}
+
s
-
\frac{iRk}{2r_h};
\Delta;z
\right),
\label{eq:KBTZ_explicit}
\end{equation}
and \( s=\pm\frac{iR^2\omega}{2r_h}\). 
This matches the standard form, which can be understood in the following way. At \(a=1\), the local Heun solution obeys the identity
\begin{equation}
\operatorname{HeunG}
\left(
1,q,\alpha,\beta,\gamma,\delta;z
\right)
=
(1-z)^s
{}_2F_1
\left(
\alpha+s,\beta+s;\gamma;z
\right),
\label{eq:Heun_to_hyper_identity}
\end{equation}
provided
\begin{equation}
q
=
\alpha\beta
+
s(\alpha+\beta-\gamma)
+
s^2.
\label{eq:Heun_identity_condition}
\end{equation}
In the present case $\gamma=\Delta$ and using \eqref{eq:alpha_beta_def}, we have
\begin{align}
\alpha+\beta
&=
\left(
\frac{\Delta}{2}
+
\frac{iRk}{2r_h}
\right)
+
\left(
\frac{\Delta}{2}
-
\frac{iRk}{2r_h}
\right)
=
\Delta\,.
\label{eq:alpha_plus_beta}
\end{align}
Hence $\alpha+\beta-\gamma=0$.
The identity condition \eqref{eq:Heun_identity_condition} therefore reduces to
\begin{equation}
q=\alpha\beta+s^2.
\label{eq:identity_condition_reduced}
\end{equation}
Demanding \(q=q_{\rm BTZ}\), this sets the value of $s$ once we plug in our results for $\alpha,\beta$. In particular using
\begin{equation}
	\alpha\beta=\left(\frac{\Delta}{2}+\frac{iRk}{2r_h}\right)\left(\frac{\Delta}{2}-\frac{iRk}{2r_h}\right)=\frac{\Delta^2}{4}+\frac{R^2k^2}{4r_h^2}\,,
\label{eq:alpha_beta_product}
\end{equation}
we get 
\begin{equation}
s^2=q_{\rm BTZ}-\alpha\beta=\left(\frac{\Delta^2}{4}+\frac{R^2k^2}{4r_h^2}-\frac{R^4\omega^2}{4r_h^2}\right)-\left(\frac{\Delta^2}{4}+\frac{R^2k^2}{4r_h^2}\right)=-\frac{R^4\omega^2}{4r_h^2}\,.
\label{eq:s_square_def}
\end{equation}
This gives the above-mentioned value of $s$. The leading BTZ kernel is therefore
\begin{equation}
K_{\rm BTZ}(z)
=\frac{1}{2\pi}\sum_{k\in\mathbb Z}
\int
\frac{d\omega}{2\pi}
e^{-i\omega (t-t')}
e^{ik(\varphi-\varphi')}
z^{\Delta/2}
(1-z)^s
{}_2F_1
\left(
\alpha+s,\beta+s;\Delta;z
\right)\,.
\label{eq:KBTZ_alpha_beta_form}
\end{equation}
The first order corrected kernel can be computed to be
\begin{align}
K_{DS}^{(1)}(z)
=\frac{1}{2\pi}\sum_{k\in\mathbb Z}
\int
\frac{d\omega}{2\pi}
e^{-i\omega (t-t')}
e^{ik(\varphi-\varphi')}
z^{\Delta/2}
&\left[
\partial_a
+
\left(
q_{\rm BTZ}
-
\frac{\Delta}{4}
\right)
\partial_q
\right]\nonumber\\
&\times \,\text{HeunG}
\left(
a,q,
\frac{\Delta}{2}
+
\frac{iRk}{2r_h},
\frac{\Delta}{2}
-
\frac{iRk}{2r_h},
\Delta,\frac12;z
\right)
\bigg|_{a=1,q=q_{\rm BTZ}}\,.
\label{eq:Kcorr_explicit}
\end{align}

Plugging the series expanded smearing Kernel back into the bulk reconstruction formula, rewriting $\varepsilon$ in terms of $\lambda$, the small-\(\lambda\) field expansion takes the form (here the subscript $R$ denotes the right-exterior field)
\begin{equation}
\Phi_R^{\rm DS}(t,z,\varphi)
=
\Phi_R^{\rm BTZ}(t,z,\varphi)
+
\frac{\lambda^2}{r_h^2}
\delta\Phi_R(t,z,\varphi)
+
O\left(\frac{\lambda^4}{r_h^4}\right)\,.
\label{eq:PhiDS_final_split}
\end{equation}
Explicitly, the BTZ piece is
\begin{align}
\Phi_R^{\rm BTZ}(t,z,\varphi)
=\frac{1}{2\pi}\int dt' \int d\varphi'
\sum_{k\in\mathbb Z}
\int
\frac{d\omega}{2\pi}
e^{-i\omega (t-t')}
e^{ik(\varphi-\varphi')}
z^{\Delta/2}
(1-z)^s \nonumber\\
{}_2F_1
\left(
\frac{\Delta}{2}
+
s
+
\frac{iRk}{2r_h},
\frac{\Delta}{2}
+
s
-
\frac{iRk}{2r_h};
\Delta;z
\right)
\mathcal O(t',\varphi')
+
\mathrm{h.c.}\,,
\label{eq:PhiBTZ_explicit}
\end{align}
with
\begin{equation}
s=\pm\frac{iR^2\omega}{2r_h}\,.
\label{eq:s_def_final}
\end{equation}
This matches with the standard HKLL BTZ field (see e.g. equation (62) of \cite{Hamilton:2006fh}). 

On the other hand, the first order finite-throat correction is given by 
\begin{eqnarray}
\delta\Phi_R(t,z,\varphi)
=\frac{1}{2\pi}\int dt'\int d\varphi'
\sum_{k\in\mathbb Z}
\int
\frac{d\omega}{2\pi}
e^{-i\omega (t-t')}
e^{ik(\varphi-\varphi')}
z^{\Delta/2}
\left[
\partial_a
+
\left(
q_{\rm BTZ}
-
\frac{\Delta}{4}
\right)
\partial_q
\right]
\nonumber\\
\operatorname{HeunG}
\left(
a,q,
\frac{\Delta}{2}
+
\frac{iRk}{2r_h},
\frac{\Delta}{2}
-
\frac{iRk}{2r_h},
\Delta,\frac12;z
\right)
\bigg|_{a=1,q=q_{\rm BTZ}}
\mathcal O(t',\varphi')
+
\mathrm{h.c.}\,,
\label{eq:deltaPhi_explicit}
\end{eqnarray}
where
\begin{equation}
q_{\rm BTZ}
=
\frac{\Delta^2}{4}
+
\frac{R^2k^2}{4r_h^2}
-
\frac{R^4\omega^2}{4r_h^2}\,.
\label{eq:qBTZ_final}
\end{equation}
Thus the scalar field admits the perturbative split
\begin{equation}
\Phi_R^{\rm DS}
=
\Phi_R^{\rm BTZ}
+
\frac{\lambda^2}{r_h^2}
\delta\Phi_R
+
O\left(\frac{\lambda^4}{r_h^4}\right)\,,
\label{eq:DS_BTZ_correction_summary}
\end{equation}
with the leading piece being the local, semiclassical bulk field in BTZ background. The appearance of the first correction is generated by the displacement of the Heun singularity \(a=z_s\) away from \(1\) together with the induced shift of the accessory parameter \(q_+\), and gives rise to an order $\lambda^2$ deformation to the BTZ reconstruction.

\section{Near-throat dynamics in DS$_3$ wormhole}
\label{subsec:near_throat_3D}

In the wormhole geometry \(\lambda\neq 0\) removes the horizon and replaces it with a smooth throat at \(r=r_h\). In the spirit of section \ref{sec:AdS2_DSW_cutoff}, the relevant question is how the throat parameter controls the local behavior of modes and ultimately the discreteness of the spectrum.
As mentioned before, for this analysis, one switches to the near-throat coordinate $\rho$, in terms of which the scalar equation takes the form of \eqref{eq:exact_rho_equation}. We have rewritten that equation below for reader's convenience:
\begin{equation}
\frac{1}{R^2}\frac{1}{\sqrt{\rho^2+\lambda^2}}
\frac{d}{d\rho}\!\left(
(r_h^2+\rho^2)\sqrt{\rho^2+\lambda^2}\,\frac{d\mathcal{R}}{d\rho}
\right)
+
\left(
\frac{R^2\omega^2}{\rho^2+\lambda^2}
-\frac{k^2}{r_h^2+\rho^2}
-m^2
\right)\mathcal{R}
=0\,.
\label{eq:exact_rho_equation_app}
\end{equation}
In the near-throat limit \(\rho\ll r_h\), the equation boils down to 
\begin{equation}
\mathcal{R}_{\rho\rho}
+\frac{\rho}{\rho^2+\lambda^2}\mathcal{R}_\rho
+
\left(
\frac{R^4\omega^2}{r_h^2(\rho^2+\lambda^2)}
-\frac{R^2}{r_h^2}\left(m^2+\frac{k^2}{r_h^2}\right)
\right)\mathcal{R}=0\,,
\qquad (\rho\ll r_h)\,.
\label{eq:near_throat_master_compact}
\end{equation}

\noindent
\emph{\textbf{(i) Sub-regime I: deep-core} \(\rho\ll \lambda\)}
\label{subsubsec:core_rho_ll_lambda}

If \(\rho\ll \lambda\), then we can approximate
\begin{equation}
\frac{\rho}{\rho^2+\lambda^2}\simeq \frac{\rho}{\lambda^2}\,.
\label{eq:rho_over_sum_core}
\end{equation}
In the strict \(\rho\to 0\) limit the term \((\rho/\lambda^2)\,\mathcal{R}_\rho\) is subleading compared to \(\mathcal{R}_{\rho\rho}\). So \eqref{eq:near_throat_master_compact} reduces at leading order to
\begin{equation}
\mathcal{R}_{\rho\rho}+\kappa_0^2\,\mathcal{R}=0\,,
\label{eq:core_constant_coeff}
\end{equation}
where
\begin{equation}
\kappa_0^2
\equiv
\frac{R^4\omega^2}{r_h^2\lambda^2}
-\frac{R^2}{r_h^2}\left(m^2+\frac{k^2}{r_h^2}\right)\,.
\label{eq:kappa0_def}
\end{equation}
Hence, the deep-core solutions are elementary. 
The solutions are 
\begin{equation}
\mathcal{R}(\rho)
=
C_1\cos(\kappa_0\rho)+C_2\sin(\kappa_0\rho)\,,
\qquad (\kappa_0^2>0\,,\;\rho\ll \lambda\ll r_h)\,,
\label{eq:core_solution}
\end{equation}
with the hyperbolic replacements if \(\kappa_0^2<0\).
\\

\noindent
\emph{\textbf{(ii) Sub-regime II: overlap} \(\lambda\ll \rho\ll r_h\)}
\label{subsubsec:overlap_bessel}

If \(\lambda\ll \rho\ll r_h\), then \(\rho^2+\lambda^2\simeq \rho^2\) and
\begin{equation}
\frac{\rho}{\rho^2+\lambda^2}\simeq \frac{1}{\rho}\,,
\qquad
\frac{1}{\rho^2+\lambda^2}\simeq \frac{1}{\rho^2}\,.
\label{eq:overlap_approxs}
\end{equation}
In this limit, after some algebra one finds
\begin{equation}
\mathcal{R}_{\rho\rho}
+\frac{1}{\rho}\mathcal{R}_\rho
+\left(\frac{\nu^2}{\rho^2}-\mu^2\right)\mathcal{R}=0\,,
\label{eq:overlap_eq_nu_mu}
\end{equation}
which have solutions in terms of the modified Bessel functions
\begin{equation}
\mathcal{R}(\rho)
=
C_1\,I_{i\nu}(\mu\rho)+C_2\,K_{i\nu}(\mu\rho),
\qquad
\nu=\frac{R^2\omega}{r_h},
\qquad
\mu^2=\frac{R^2}{r_h^2}\left(m^2+\frac{k^2}{r_h^2}\right)\,.
\label{eq:overlap_solution}
\end{equation}

\noindent
\emph{\textbf{Special case:} \(\mu=0\)}

If \(m=0\) and \(k=0\), then \(\mu=0\) and \eqref{eq:overlap_eq_nu_mu} reduces to
\begin{equation}
\mathcal{R}_{\rho\rho}+\frac{1}{\rho}\mathcal{R}_\rho+\frac{\nu^2}{\rho^2}\mathcal{R}=0\,.
\label{eq:mu0_eq}
\end{equation}
It has power-law solutions are \(\mathcal{R}\sim \rho^{\pm i\nu}\).

\section{WKB data: turning point and symmetric quantization}
\label{subsec:WKB_data}

We briefly collect the numerical ingredients used for the WKB spectra and spectral form factors that we have used in section \ref{sec:numerics_SFF_3D}. Starting from the Schrödinger-form radial equation
\begin{equation}
    \frac{d^2\varphi}{dr_*^2}
    +
    \left(\omega^2 - V_{\rm eff}(r)\right)\varphi
    =
    0\,,
    \label{eq:schrodinger_wkb_appendix}
\end{equation}
with $V_{\rm eff}(r)$ given in \eqref{eq:Veff_full_rewrite}, the non-zero throat parameter $\lambda$ makes the geometry smooth at $r=r_h$. For fixed angular momentum $k$, the outer turning point $r_t>r_h$ is determined by
\begin{equation}
    \omega^2 = V_{\rm eff}(r_t)\,,
    \qquad r_t>r_h
    \label{eq:turning_point_appendix}
\end{equation}
with an identical turning point on the other side of the throat. The corresponding Bohr--Sommerfeld condition is therefore\footnote{Alternatively we can perform this analysis using the tortoise coordinate $x$ introduced in \eqref{eq:xu_def}, in which the throat is the midpoint of a symmetric one-dimensional potential well. This symmetricity is what gives rise to the factor of 2 in \eqref{eq:bohr_sommerfeld_appendix}.}
\begin{equation}
    2\int_{r_h}^{r_t}
    \frac{R^2\,dr}{\alpha(r)\beta(r)}
    \sqrt{\omega_n^2 - V_{\rm eff}(r)}
    =
    \pi\left(n+\frac12\right)\,,
    \qquad n\in \mathbb{Z}_{\geq 0}
    \label{eq:bohr_sommerfeld_appendix}
\end{equation}
where
\begin{equation}
    \alpha(r)=\sqrt{r^2-r_h^2+\lambda^2}
    \qquad \text{and}\qquad
    \beta(r)=\sqrt{r^2-r_h^2}\,.
    \label{eq:alpha_beta_appendix}
\end{equation}
This is the quantization condition used to generate the discrete WKB levels $\omega_{n,k}$.

Numerically, for each fixed $k$ and $n$, we solve the Bohr--Sommerfeld condition by bisection in $\omega$. At every trial value of $\omega$, the corresponding turning point $r_t(\omega)$ is determined by solving $\omega^2-V_{\rm eff}(r)=0$ for $r>r_h$. This gives the WKB action $I(\omega,k)$, and the outer bisection is continued until
\begin{equation}
    I(\omega,k)-\pi\left(n+\frac12\right)=0 \,.
    \label{eq:wkb_bisection_condition}
\end{equation}
Thus $r_t$ is not an independent input; it is recomputed as a function of each trial $\omega$. The square-root behavior near the endpoint is regulated numerically by writing
\begin{equation}
    r(u)=r_h+(r_t-r_h)u^2\,,
    \qquad 0\leq u\leq 1 \,.
    \label{eq:endpoint_regularization}
\end{equation}
The WKB action is then evaluated as
\begin{equation}
    I(\omega,k)
    =
    2\int_0^1 du\,
    2(r_t-r_h)u\,
    \frac{R^2}{\alpha(r(u))\beta(r(u))}
    \sqrt{\omega^2 - V_{\rm eff}(r(u))}\,.
    \label{eq:wkb_action_regularized}
\end{equation}
The frequencies are obtained within the finite windows of $n$ and $k$ used in the numerical plots.

Given the resulting discrete spectrum, the spectral form factor is computed from
\begin{equation}
g(\beta_E,t)
    =
    \frac{|Z(\beta_E,t)|^2}{|Z(\beta_E,0)|^2}
    \qquad \text{with}\qquad
    Z(\beta_E,t)
    =
    \sum_a e^{-(\beta_E+it)\omega_a}\,,
\label{eq:sff_partition_appendix}
\end{equation}
where $a$ denotes the combined label $(n,k)$ over the selected levels. At $\beta_E=0$, this reduces to
\begin{equation}
    g(it)=
    \frac{\left|\sum_a e^{-it\omega_a}\right|^2}
    {\left|\sum_a 1\right|^2}\,.
    \label{eq:sff_beta_zero_appendix}
\end{equation}

For the $\lambda$-averaged diagnostic, one samples a narrow distribution of throat parameters $\{\lambda_i\}_{i=1}^{N}$ around a mean value $\lambda_0$ (taken from a Gaussian ensemble), recomputes the spectrum for each $\lambda_i$, and averages the normalized SFF
\begin{equation}
    \left\langle g(\beta_E,t)\right\rangle_\lambda
    =
    \frac{1}{N}\sum_{i=1}^{N}
    \frac{|Z_{\lambda_i}(\beta_E,t)|^2}
    {|Z_{\lambda_i}(\beta_E,0)|^2}\,.
    \label{eq:lambda_averaged_sff_appendix}
\end{equation}
This averaging suppresses the non-universal oscillations of a fixed spectrum and exposes the coarse dip--ramp--plateau structure discussed in section \ref{sec:numerics_SFF_3D}.


\bibliographystyle{utphys}
\bibliography{ref}
\end{document}